\documentclass[11pt]{iopart}
\usepackage[english]{babel}
\pdfoutput=1
\usepackage{wasysym}
\usepackage{booktabs}
\usepackage{amssymb}
\usepackage{amsbsy}
\usepackage{verbatim}
\usepackage{graphicx}
\usepackage{epstopdf}
\usepackage{color}
\usepackage{sidecap}
\usepackage{bm}

\usepackage[colorlinks,bookmarks=false,citecolor=blue,linkcolor=red,urlcolor=blue]{hyperref}

\catcode`@=11 
\renewcommand\tableofcontents{%
  \section*{\contentsname}%
  \@starttoc{toc}%
}
\catcode`@=12

\newcommand{\ket}[1]{\left|{#1}\right.\rangle}

\def\be{\begin{equation}}
\def\ee{\end{equation}}

\def\nm{\newmoon}
\def\fm{\fullmoon}
\def\T{\rule{0pt}{.6cm}}
\def\B{\rule[-.4cm]{0pt}{0pt}}

\begin{document}

\setlength{\parindent}{0pt}

\title{Bethe ansatz description of edge-localization in the open-boundary 
XXZ spin chain}

\author{Vincenzo Alba$^1$, Kush Saha$^{2,3}$, and Masudul Haque$^3$}

\address{$^1$ Department of Physics and Arnold Sommerfeld
Center for Theoretical Physics, Ludwig-Maximilians-Universit\"at
M\"unchen, D-80333 M\"unchen, Germany}

\address{$^2$Theoretical Physics Department, Indian Association for
the Cultivation of Science, Kolkata-700032, India.}

\address{$^3$
 Max-Planck-Institut f\"{u}r Physik komplexer Systeme,
  N\"{o}thnitzer Stra{\ss}e 38, D-01187 Dresden, Germany}




\date{\today}

\begin{abstract} 

At large values of the anisotropy $\Delta$, the open-boundary Heisenberg spin-$\frac{1}{2}$ chain
has eigenstates displaying localization at the edges.  We present a Bethe ansatz description of this
`edge-locking' phenomenon in the entire $\Delta>1$ region.  We focus  on the simplest spin sectors,
namely the highly polarized sectors with only one or two overturned spins, i.e., one-particle and
two-particle sectors. 

Edge-locking is associated with pure imaginary solutions of the Bethe equations, which are not
commonly encountered in periodic chains.  In the one-particle case, at large $\Delta$ there are two
eigenstates with imaginary Bethe momenta, related to localization at the two edges.  For any finite
chain size, one of the two solutions become real as $\Delta$ is lowered below a certain value.  

For two particles, a richer scenario is observed, with eigenstates having the possibility of both
particles locked on the same or different edge, one locked and the other free, and both free either
as single magnons or as bound composites corresponding to `string' solutions.  For finite chains,
some of the edge-locked spins get delocalized at certain values of $\Delta$ (`exceptional points'),
corresponding to imaginary solutions becoming real.  We characterize these phenomena thoroughly by
providing analytic expansions of the Bethe momenta for large chains, large anisotropy $\Delta$, and
near the exceptional points.  In the large-chain limit all the exceptional points coalesce at the
isotropic point ($\Delta=1$) and edge-locking becomes stable in the whole $\Delta>1$ region.

\end{abstract}

\maketitle

\section{Introduction}
\label{intro}


\paragraph*{General context.}

The effects on many-body systems exerted by localized features such as edges and impurities have long
been a central theme of condensed matter physics.  Kondo physics~\cite{kondo-1964} and Anderson
orthogonality catastrophe~\cite{anderson-1967} are among the most celebrated examples of complex
physics caused by single impurities.  An edge of a finite system can also be responsible for
families of effects.  Among other phenomena, an edge can bind or lock excitations or particles.
Intriguingly, in addition to single-particle binding at boundaries and edges, edge-locking can also
arise as a collective interaction-induced phenomenon, which  results in unintuitive temporal
dynamics \cite{haque-09,haque-10}. 

Edges appear naturally through the use of open boundary conditions.  Periodic boundary conditions
are, of course, far more popular due to the presence of translation symmetry and due to having
physical momentum as a good quantum number.  In particular, in the Bethe ansatz approach to
one-dimensional (1D) systems, the bulk of the literature focuses on periodic chains.  Nevertheless,
since the discovery of the exact solution of the spin-$\frac{1}{2}$ Heisenberg chain with boundary
magnetic fields~\cite{alcaraz-87,sklyanin-88}, integrable models have provided a rich playground to
investigate edge-related physics in 1D.  Example topics studied are edge bound states in integrable
field theories~\cite{skorik-95,kapustin-96,chihiro-12,fendley-94,leclair-95,asakawa-97}, Kondo-like
behaviors in spin chains~\cite{desa-95,wang-96,zvyagin-97,schlott-97,wang-97}, Friedel
oscillations~\cite{asakawa-98}, Anderson orthogonality
catastrophe~\cite{bedur-97,frahm-98,bortz-05}, with also potential applications in quantum computing
devices~\cite{dykman-03,santos-03}.

In this work, we focus on the anisotropic Heisenberg (XXZ) chain, and examine structures that appear
in the spectra due to the presence of open boundary conditions (edges), in particular those
eigenstates whose spatial forms are dictated by the edge.  In these eigenstates, one or more
``particles'' (overturned spins) are localized or locked at the edges.  In the sectors we look at
(one or two particles), most of these features are physically simple and show up as prominent band
structures in the spectrum, as we show below in Figure \ref{fig_spectra_periodic_vs_open}.  However,
a Bethe ansatz description has been lacking in the literature to the best of our knowledge, despite
the relevant Bethe equations having been available since the work of
Refs.~\cite{alcaraz-87,sklyanin-88}.


\paragraph*{The open-boundary Heisenberg XXZ chain.}

The open-boundary anisotropic spin-$1\over 2$ Heisenberg XXZ chain of $L$ interacting spins is
described by the Hamiltonian
\begin{equation}
{\mathcal H}=\frac{1}{2}\sum_{i=1}^{L-1}(S_i^{+}S_{i+1}^{-}+
S_{i}^-S_{i+1}^+)+\Delta\sum_{i=1}^{L-1}S_{i}^zS_{i+1}^z
\label{xxz_ham}
\end{equation}
Here $S_i^{+,-,z}$ are given as $S_i^{\pm}\equiv(\sigma_i^x\pm i \sigma_i^y)/2,S_i^z\equiv
\sigma_i^z/2$, with $\sigma_i^{x,y,z}$ the Pauli matrices, and $\Delta$ is the anisotropy.  The
sites $i=1$ and $i=L$ are endpoints and are only coupled to one neighbor each.  The Hilbert space
of~\eref{xxz_ham} is spanned by $2^L$ basis states, which are conveniently generated starting from
the fully polarized eigenstate $\left|0\right.\rangle \equiv
\left|\uparrow\uparrow\uparrow\cdots\uparrow\right.\rangle$ with all the spins up, and overturning
$M$ of the spins, with $M\in[0,L]$.  As conventional in the Bethe ansatz literature, we refer to
overturned spins as ``particles''.

\begin{figure}[t]
\begin{center}
\includegraphics[width=.6\textwidth]{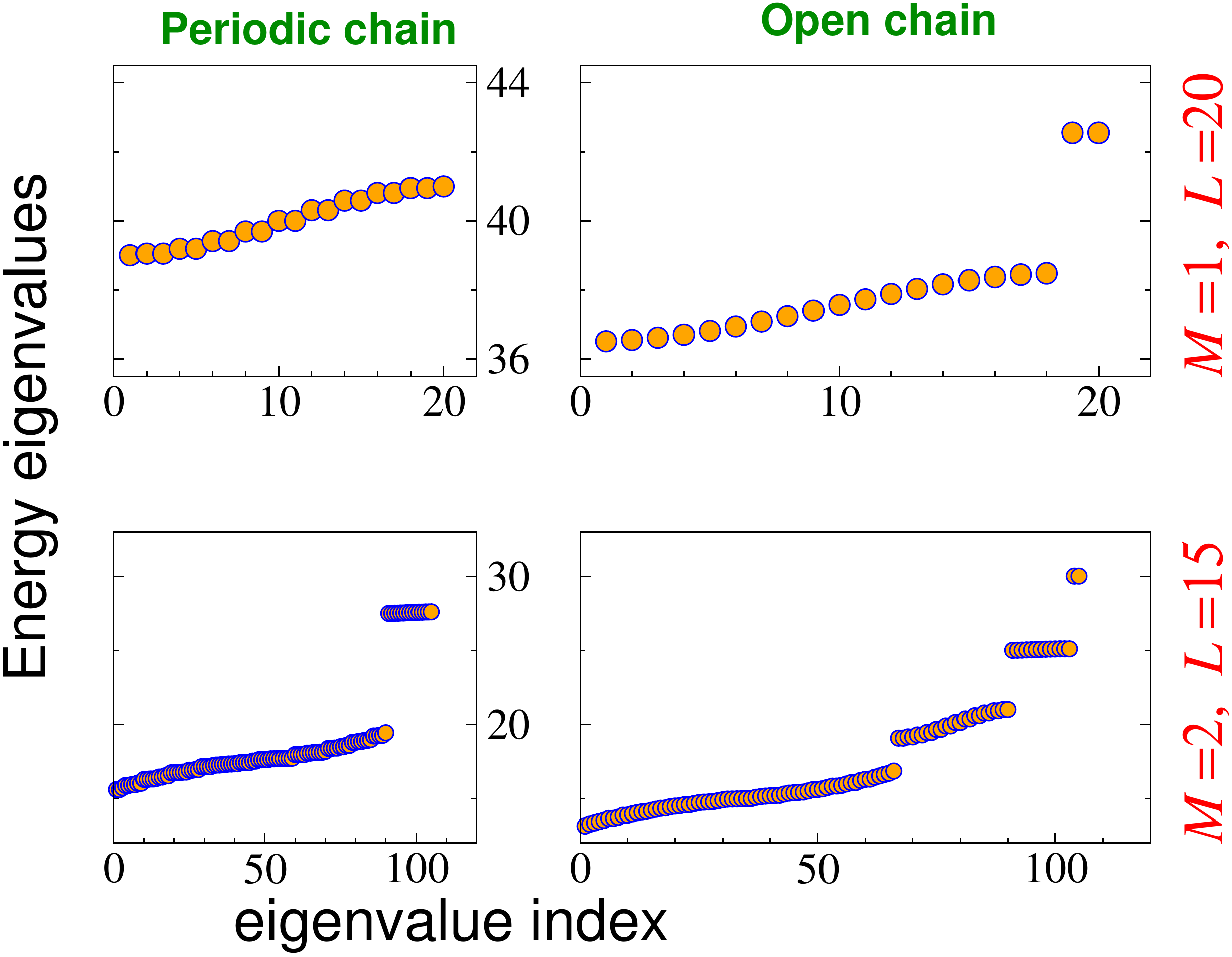}
\end{center}
\caption{Comparison of the energy spectra of periodic and open chains, $M=1$ and $M=2$ sectors,
  obtained by numerical diagonalization of the Hamiltonian at $\Delta=10$.  Energy eigenvalues are
  ordered in increasing algebraic order.  }
\label{fig_spectra_periodic_vs_open}
\end{figure}

We restrict to the highly polarized sectors of \eref{xxz_ham} with only one ($M=1$) and two ($M=2$)
particles.  The spectral effect of open boundary conditions in these sectors is shown in Figure
\ref{fig_spectra_periodic_vs_open}, where periodic and open boundary conditions are compared through
numerical diagonalization. The extra structures in the open-chain cases clearly represent edge
physics: for example, the top two eigenstates correspond to the one ($M=1$) or two ($M=2$) particles
localized at the two edges of the chain.

In this work, through explicit consideration of the full spectrum of~\eref{xxz_ham} via the Bethe
ansatz formalism, we provide a complete classification of edge-locking behavior in the whole region
$1<\Delta<\infty$ in the $M=1,2$ sectors.

\paragraph*{Outline of main results.}

We show that edge-locking is clearly reflected in the nature of the solutions of the Bethe equations
(Bethe momenta): while locked particles correspond to pure imaginary momenta, extended or
magnon-like behavior is signaled by real solutions.

As a consequence, edge-locking provides a useful physical criterion for classifying the Bethe momenta.  In
fact, a byproduct of our analysis is a complete scrutiny of the {\it full} set of solutions of Bethe
equations in the $M=1,2$ sectors at $1<\Delta<\infty$.  This is similar to what has been done
in~\cite{karbach-1997,essler-92} for the isotropic ($\Delta=1$) Heisenberg model with periodic
boundary conditions.

In the $M=1$ sector, there is a single Bethe momentum describing the eigenstates and only two types
of eigenstates are possible.  The Bethe momentum is either real, corresponding to a spatially
extended particle, or purely imaginary, corresponding to the particle being edge-locked.  Due to
reflection symmetry, the eigenstates with edge-locking are linear combinations of configurations
with locking at the left edge and at the right edge.  There are clearly two such eigenstates at
large $\Delta$.  Remarkably, for any finite chain length $L$, as $\Delta$ is lowered there is a
value of $\Delta$ where one of these eigenstates gets delocalized, and the corresponding Bethe
momentum becomes real instead of imaginary.  Motivated by real$\leftrightarrow$imaginary
transformation effects in the literature on non-hermitian matrices, we refer to such special $\Delta$
values as ``exceptional points''.

In the $M=2$ sector a richer scenario arises: eigenstates can have both particles edge-locked (fully
edge-locked states), or no edge-locking (both particles  spatially extended or magnon-like), or
have one particle edge-localized and the other spatially extended.  These classes correspond to
different classes of Bethe momenta: both momenta imaginary, both real or a complex conjugate pair,
one real and one imaginary.
At large $\Delta$, using geometric arguments, one can find the numbers of different classes of
eigenstates, as simple functions of $L$ (Section \ref{two_part}).  At smaller $\Delta$, there are
several series of exceptional points at which imaginary momenta become real, and eigenstates lose
part or all of their edge-locked nature, i.e, particles ``delocalize''.

In both $M=1,2$ sectors, two fully edge-locked eigenstates form a doublet of quasi-degenerate energy
levels, which is separated from the rest of the spectrum by a gap $\sim\Delta$, as can be seen in
Figure \ref{fig_spectra_periodic_vs_open}.  The energy splitting within the doublet vanishes
exponentially with the system size.  In the $M=2$ case, these are eigenstates where the two
particles are localized on the same edge.  A third fully edge-locked state has one particle
localized at each edge; this state is not spectrally separated but has an intriguingly simple
structure (Section \ref{edge_eig_2p}).

In this work we throughly characterize the natures of the eigenstates in the Bethe ansatz language,
by providing the numbers of different classes of eigenstates at different $\Delta$ values, the
locations of the exceptional points $\Delta_e$, analytic expressions for the Bethe momenta at large
$\Delta$ and large $L$ and in the vicinity of the exceptional points, etc.  The values of the
exceptional points are described by a set of two coupled equations that we provide explicitly.  An
intriguing feature is that in the limit $L\to\infty$ all the exceptional points coalesce at the
isotropic point $\Delta=1$, signaling that edge-locked particles become stable in the whole region
$1<\Delta<\infty$.  Finally, we discuss $M=2$ eigenstates where the particles are extended but
mutually bound, which correspond to complex conjugate pairs of Bethe momenta, and are closely
analogous to ``2-strings'' well-known from the periodic chain.  These eigenstates are
found to be stable in the whole $\Delta>1$ region, i.e. no unbinding of bound states or locking at
the boundary is observed.

\paragraph*{Organization of this Article.}

In section \ref{sec_1_BA} we outline the Bethe ansatz formulation for the open XXZ chain, following
\cite{alcaraz-87}.  Section \ref{one_p} describes the one-particle ($M=1$) sector, characterizing
edge-locked and extended states and the exceptional point below which one of the edge-locked states
becomes extended.  The next six sections, \ref{two_part} to \ref{sec_2p_spectrum}, detail the
two-particle ($M=2$) sector.  We start this discussion in Section \ref{two_part} with an outline of
the different types of solutions expected at large $\Delta$ from physical expectations of
edge-locking, and then describe the different types of solutions (real+real, imaginary+imaginary,
real+imaginary, complex conjugate pairs) in the next few sections, respectively Sections
\ref{sec_twoparticle_bothreal}, \ref{pure_im}, \ref{re_im}, \ref{string}.  We end the discussion of
the $M=2$ sector in Section \ref{sec_2p_spectrum} with an overview of the spectrum, more detailed
than that provided in Figure \ref{fig_spectra_periodic_vs_open}.  Section \ref{conclusions}
concludes the article.





\section{Bethe ansatz approach}
\label{sec_1_BA}

We start with setting up the notation for the Bethe ansatz approach~\cite{alcaraz-87, sklyanin-88,
  bethe-31, kor-book} for the XXZ spin chain with open boundary conditions. First, since the total
magnetization $S_T^z\equiv\sum_iS^z_i=L/2-M$, $M$ being the number of particles, is a conserved
quantity, it can be used to label the eigenstates of~\eref{xxz_ham}. We denote as $|\Psi_M\rangle$
an eigenstate of~\eref{xxz_ham} in the sector with $M$ particles. This, for any $\Delta$, can be
written in general as

\begin{equation}
|\Psi_M\rangle=\sum\limits_{1\le j_1<j_2<\dots<j_M\le L} A_M (j_1,j_2,j_3...,
j_M)|j_1,j_2,j_3....j_M\rangle
\label{B_ansatz}
\end{equation}

where the sum is over the positions $1\le j_n\le L$ ($n=1,2,\dots,M$) of the particles 
in the chain, while $A_M(\{j_n\})$ is the amplitude of the eigenstate component 
with particles at positions $j_1,j_2,\dots,j_M$. In the Bethe ansatz approach one 
rewrites~\eref{B_ansatz} as

\begin{equation}
\fl
A_M(j_1,j_2,\dots,j_M) ~=~ 
\sum\limits_{\bar {\mathcal P}}  (-1)^{\epsilon_{\bar{\mathcal P}}} \,  
B_M(k_{\bar{\mathcal P}_1},k_{\bar{\mathcal P}_2},
\dots,k_{\bar{\mathcal P}_M})  \, 
e^{i(k_{\bar{\mathcal P}_1}j_1+k_{
\bar{\mathcal P}_2}j_2+\dots+k_{\bar{\mathcal P}_M}j_M)}
\label{A_amplitude}
\end{equation}

Here the sum is over all the permutations and (arbitrary number of) reflections
$k_i\to -k_i$ of the so-called Bethe momenta $k_1,k_2,\dots, k_M$, while 
$\epsilon_{\bar{\mathcal P}}$ is given as $\epsilon_{\bar{\mathcal P}}
\equiv \epsilon_{\mathcal P}+ \varepsilon$, 
with $\epsilon_{\mathcal P}$ the sign of the permutation $\mathcal P$ and $\varepsilon$ the number
of reversed momenta. The Bethe momenta $k_\ell$ with $\ell=1,2,\dots,M$ are solutions of the non
linear set of equations (Bethe equations)

\begin{equation}
\fl e^{2i(L+1)k_\ell}=\frac{(1-\Delta e^{ik_\ell})^2}{(1-\Delta e^{-ik_\ell})^2}
\prod_{j=1,j\ne \ell}^M\frac{(1+e^{i(k_j-k_\ell)}-2\Delta e^{ik_j})
(1+e^{i(k_\ell+k_j)}-2\Delta e^{ik_\ell})}
{(1+e^{i(k_\ell+k_j)}-2\Delta e^{ik_j})(1+e^{i(k_j-k_\ell)}-2\Delta e^{-ik_\ell})}
\label{B_equations}
\end{equation}

In terms of $\{k_\ell\}$ the amplitude $B_M$ is given as

\begin{equation}
\fl B_M ~=~ \prod\limits_{j=1}^M(1-\Delta e^{ik_j})e^{-i(L+1)k_j}
\prod\limits_{1\le j<\ell\le M}(1+e^{i(k_\ell-k_j)}-2\Delta e^{ik_\ell})
(1+e^{i(k_j+k_\ell)}-2\Delta e^{ik_j})e^{-ik_\ell}
\label{B_amplitude}
\end{equation}

For each set of solutions of~\eref{B_equations} one has that~\eref{B_ansatz} 
is an eigenstate of~\eref{xxz_ham}. The corresponding energy is given as 

\begin{equation}
E=\sum\limits_{i=1}^M\cos k_i+E_0(L,M),\qquad E_0(L,M)\equiv
\Big(\frac{L-1}{4}-M\Big)\Delta
\label{B_energy}
\end{equation}

Note that the amplitude $A_M(\{k_\ell\})$ is identically zero for $k_{\ell}=0$ and $k_{\ell}=\pi$,
implying that these Bethe momenta, although formally solutions of the Eqs.~\eref{B_equations}, are
not allowed.

\begin{figure}[t]
\begin{center}
\includegraphics[width=.6\textwidth]{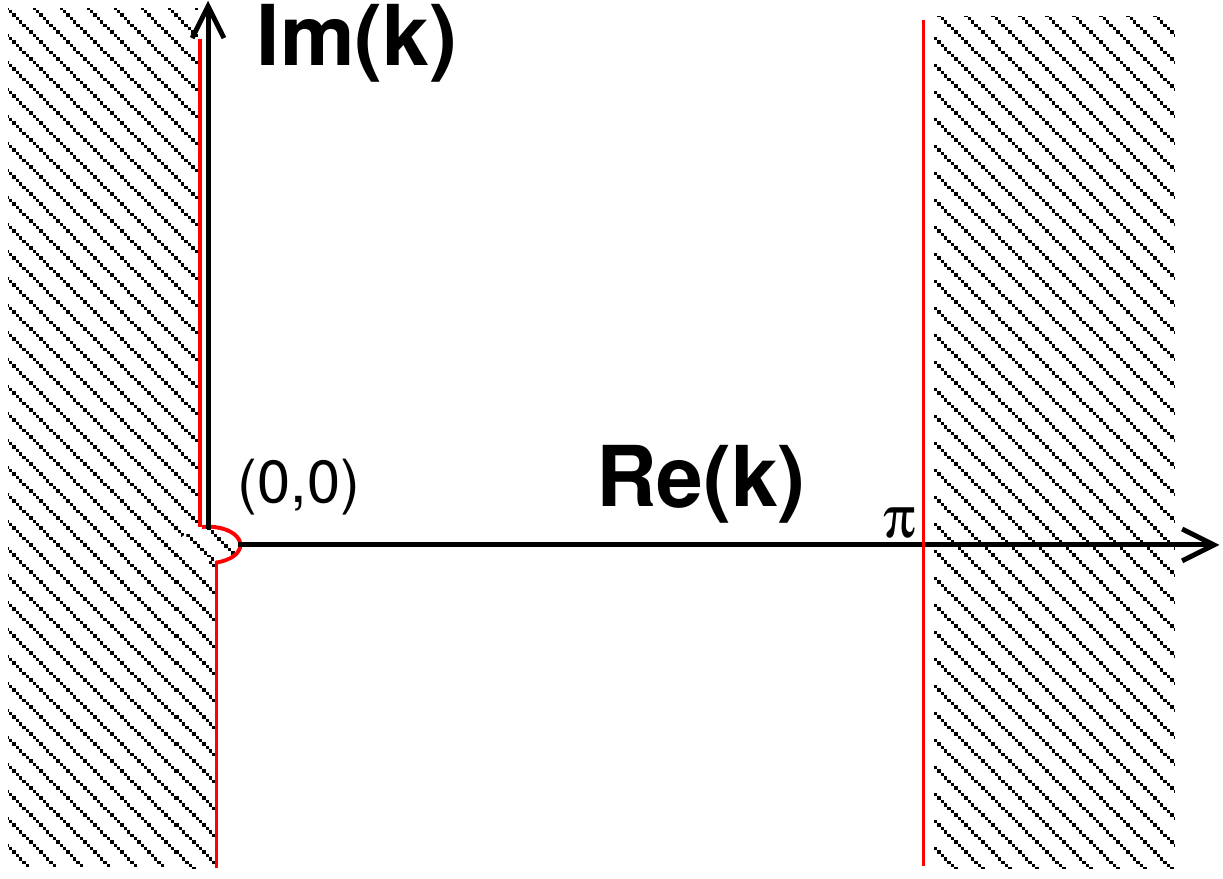}
\end{center}
\caption{ Allowed values for the Bethe momenta $k\equiv(\textrm{Re}(k), \textrm{Im}(k))$ in the
  complex plane, for the open-boundary XXZ chain. The dashed area represents the excluded regions
  The line $\textrm{Re}(k)=\pi$ is excluded.  Purely imaginary momenta with $\textrm{Im}(k)>0$ are
  allowed.  }
\label{fig_0}
\end{figure}

The amplitude~\eref{A_amplitude} is symmetric under permutations and reflections of the momenta
$k_1,k_2,\dots,k_M$. This symmetry is inherited by the Bethe momenta $\{k_\ell\}$, i.e. given the
set of solution of~\eref{B_equations} as ${\mathcal K}\equiv\{k_\ell\}$, any other set obtained by
permuting and inverting an arbitrary number of elements of ${\mathcal K}$ is also a set of solutions
of~\eref{B_equations}.  Thus one can restrict to the positive solutions of~\eref{B_equations},
i.e. requiring $k_\ell>0\,\, \forall\,\ell$. This also implies that only half of the complex plane
is allowed for $k_\ell$.  Here we choose the right half, i.e., we impose $\textrm{Re}(k)>0$.  Note
that, for purely imaginary momenta ($\textrm{Re}(k)=0$) only half of the imaginary axis is allowed
(we impose $\textrm{Im}(k)>0$). Finally, using the invariance $k_\ell\to-k_\ell$ and the $2\pi$
periodicity of the functions appearing in~\eref{B_equations} one can also restrict to
$\textrm{Re}(k)<\pi$. The resulting region of allowed values for the Bethe momenta in the complex
plane is shown in Figure~\ref{fig_0}.

\paragraph*{Comparison with periodic case.}

For periodic boundary conditions, the Bethe equations would be given as
\begin{equation}
e^{iLk_\ell}=\prod_{j=1,j\ne \ell}^M\left[-\frac{1+e^{i(k_\ell+k_j)}-
2\Delta^{ik_\ell}}{1+e^{i(k_\ell+k_j)}-2\Delta e^{ik_j}}\right]
\, . 
\label{B_equations_per}
\end{equation}
In the periodic case the Bethe momenta are not restricted to a half-plane and can range in the interval
$\textrm{Re}(k)\in(-\pi,\pi]$, i.e., double the region shown in Figure~\ref{fig_0}.

For the same $M$ sector, the Bethe equations are more complicated in the open-boundary case [compare
  Eq.~\eref{B_equations} with Eq.~\eref{B_equations_per}]; so it is not surprising that the
open-boundary case has more structure such as the possibility of purely imaginary solutions.

\section{The one particle sector: extended versus edge-locked behavior}
\label{one_p}

In this section we will treat the one-particle sector ($M=1$).

The total number of eigenstates, i.e., solutions to the Bethe equations~\eref{B_equations}, is equal
to the chain length $L$.  Since the energy eigenvalues must be real, Eq.~\eref{B_energy} constrains
the solutions of~\eref{B_equations} to be either purely real or purely imaginary.  Real
solutions correspond to spin wave states (magnons), which are extended in the bulk of the chain,
whereas purely imaginary Bethe momenta correspond to edge-locked ones. These are eigenstates with
the particle {\it exponentially} localized at the edges of the chain.

The number of imaginary solutions (edge-locked eigenstates) depends on the anisotropy $\Delta$ and
on chain length $L$.  The situation is summarized in Figure \ref{fig_1}.  At large $\Delta$ there
are two imaginary momenta ($k_{\pm}$).  For each fixed size $L$ we find that there exists an
``exceptional'' value of the anisotropy, $\Delta=\Delta_e$, at which one of the purely imaginary
solutions passes through zero and becomes purely real.  For $\Delta\le\Delta_e$ the Bethe equations
admit only one imaginary solution.  Clearly, the number of real solutions is $L-2$ and $L-1$ for
respectively $\Delta>\Delta_e$ and $\Delta<\Delta_e$ respectively.

The imaginary solution surviving in the region $\Delta<\Delta_e$
itself vanishes at $\Delta\to1$.  This means that $\Delta=1$ is also an exceptional point, and that
pure imaginary momenta are not present at $\Delta=1$.  We show below, and have found numerically,
that $\Delta_e = (L+1)/(L-1)$, so that $\Delta_e$ decreases monotonically upon increasing the chain
size and coalesces to $\Delta_e\to 1$ in the limit $L\to\infty$.

\subsection{The magnon states}

We first consider the extended eigenstates, i.e., real 
solutions of the Bethe equations. The corresponding 
Bethe momenta are obtained by solving the equation
\begin{equation}
e^{2ik(L+1)}=\frac{(1-\Delta e^{ik})^2}{(1-\Delta e^{-ik})^2}
\label{1p_equations}
\end{equation}
%
%
As usual in the Bethe ansatz literature, we consider the Bethe equation in logarithmic form.  First
we redefine the momentum $k$ in terms of the so-called rapidity $\lambda$:
\begin{equation}
k=\pi-2\arctan \frac{\tan\lambda}{\tanh(\eta/2)}-2\pi\left\lfloor
\frac{\lambda}{\pi}+\frac{1}{2}\right\rfloor
\label{1p_momentum}
\end{equation}
where $\eta$ is related to the anisotropy $\Delta$ as $\eta\equiv\textrm{arcosh}
(\Delta)$.  The term with the floor function $\lfloor\cdot\rfloor$  is 
convenient in order to make $k$ continuous as function  of $\lambda$ in the 
interval $[0,\pi]$. Taking the logarithm on both sides in~\eref{1p_equations} one 
obtains 
\begin{equation}
\fl
\arctan \frac{\tan\lambda}{\tanh(\eta/2)}+\pi\left\lfloor\frac{\lambda}
{\pi}+\frac{1}{2}\right\rfloor=\frac{\pi J}{2(L+1)}+\frac{1}{L+1}\left[
\arctan\left(\frac{\tan2\lambda}{\tanh\eta}\right)+\left\lfloor\frac{2\lambda}
{\pi}+\frac{1}{2}\right\rfloor\right]
\label{1p_equations_log}
\end{equation}
Here  the integer $J$ ($J\in[1,L]$) is the so-called Bethe quantum number. 
Each choice of $J$ identifies, in principle, an eigenstate of~\eref{xxz_ham}.
The corresponding energy expressed in terms of $\lambda$ reads 
\begin{equation}
E=-\cos\Big(2\arctan \frac{\tan\lambda}{\tanh(\eta/2)}+2\pi
\left\lfloor\frac{\lambda}{\pi}+\frac{1}{2}\right\rfloor\Big)+E_0(L,1)
\label{1p_ener_lambda}
\end{equation}

\begin{figure}[t]
\begin{center}
\includegraphics[width=0.98\textwidth]{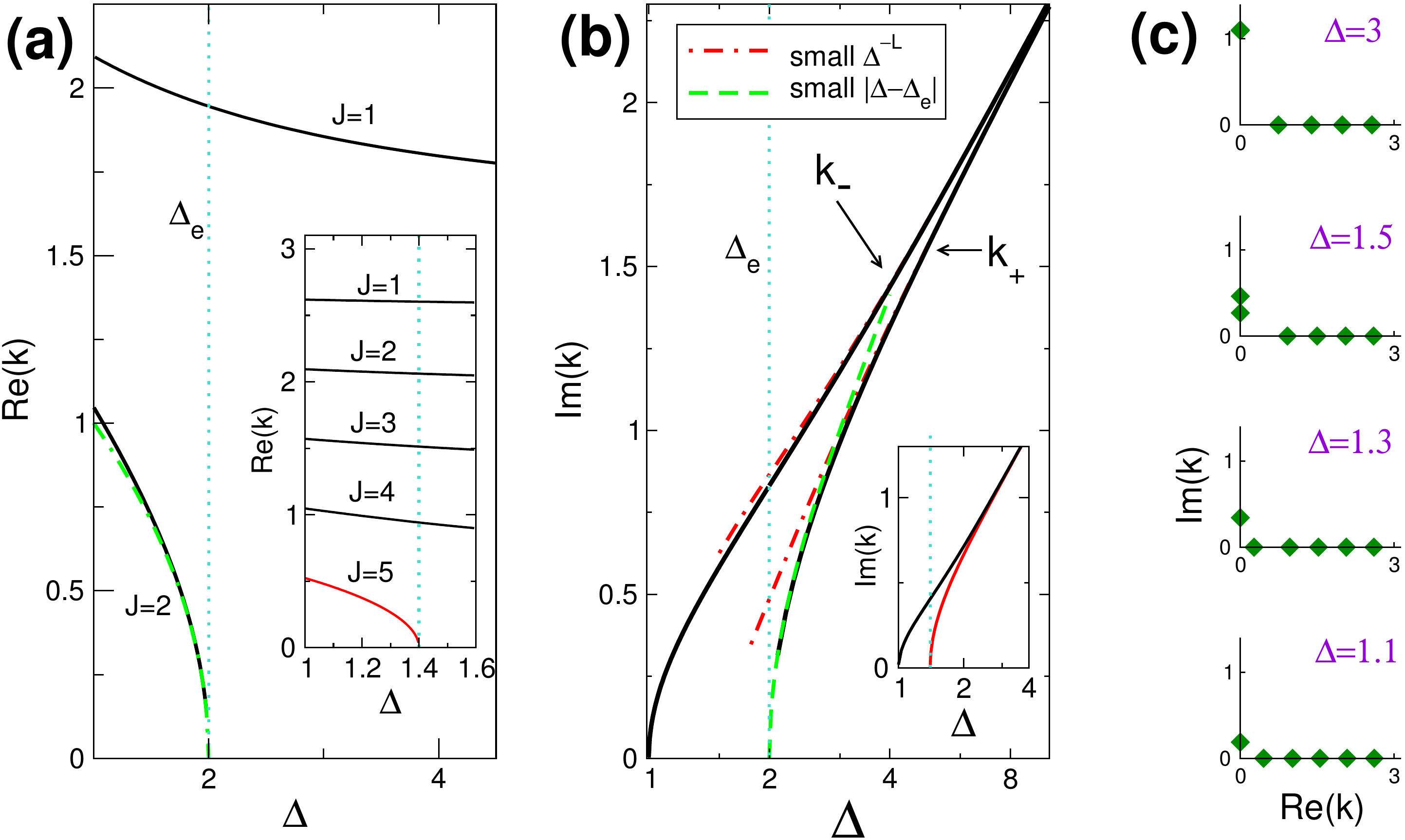}
\end{center}
\caption{ Solutions of the Bethe equation in the $M=1$ sector as a function of $\Delta$.  In
  (\textbf{a},\textbf{b}), main panels show $L=3$ and insets show $L=6$.  (\textbf{a}) Real
  solutions.  The Bethe number $J$ is noted for each solution.  There are $L-1$ real solutions for
  $\Delta < \Delta_e= \frac{L+1}{L-1}$ and $L-2$ real solutions for $\Delta>\Delta_e$.  One real
  solution ($J=L-1$) vanishes as $\Delta\to\Delta_e^-$.  The dashed dotted line is the analytic
  expression in the vicinity of $\Delta_e$.
(\textbf{b}) Imaginary solutions.  Logarithmic scale on the horizontal axis.  There are two
imaginary momenta at $\Delta>\Delta_e$; one of them ($k_+$) vanishes at $\Delta=\Delta_e$.  The dashed and
dashed-dotted lines are the analytic expansions respectively at $\Delta\to\Delta_e$ and at large
$\Delta$.
(\textbf{c}) Solutions on complex-momentum plane, $L=6$, shown for a sequence of $\Delta$ values
above and below $\Delta_e=1.4$.  Real and imaginary axes have different scales.
}
\label{fig_1}
\end{figure}

Figure \ref{fig_1}(a) shows real solutions obtained from the Bethe equation.  For $\Delta>\Delta_e$,
real rapidities and hence real momenta are found by using values $J=1,2,\dots, L-2$.  For
$\Delta<\Delta_e$, there is an extra real solution as $J=L-1$ also gives a real rapidity.  In the
Figure, the special value is $\Delta_e=$ 2(1.4) for $L=$ 3(6), consistent with the relation
$\Delta_e=(L+1)/(L-1)$ derived below.

\subsection{The edge-locked states (half-strings)}

In this section we discuss the nature of the imaginary solutions of the Bethe equations and their
evolution as function of $\Delta$ across the exceptional point $\Delta_e$.  Due to the restriction
$\textrm{Im}(k)>0$, the purely imaginary strings do not appear in complex conjugate pairs.  For this
reason we refer to them as ``half-strings''.

Figure~\ref{fig_1}(b) plots the two imaginary solutions ($k_+,k_-$) of the Bethe
equations~\eref{1p_eqns_im} for an open XXZ chain with $L=3$ as function of $\Delta$.  At
$\Delta>\Delta_e$, two imaginary momenta are present. At $\Delta\to\Delta_e$ one of the two ($k_+$)
vanishes.  This solution re-emerges on the other side of $\Delta_e$ as a real solution.  This can be
seen on the sequence of plots in Figure~\ref{fig_1}(c). 

In order to study the imaginary solutions of~\eref{1p_equations} we start redefining $i\log z\equiv
k$ with $z$ real. Since $\textrm{Im}(k)>0$ (Figure~\ref{fig_0}), we have $z>1$. The Bethe equation
is 
\begin{equation}
z^{-2L}=\frac{(z-\Delta)^2}{(1-\Delta z)^2}
\label{1p_eqns_im}
\end{equation}
Clearly the left side in Eq.~\eref{1p_eqns_im} vanishes
exponentially upon increasing $L$. To recover the same behavior
in the right side one has to impose $z=\Delta+{\mathcal O}
(e^{-L})$.  Substituting this ansatz 
into~\eref{1p_eqns_im} one finds two solutions $z_\pm$  as
\begin{equation}
z_{\pm}=\Delta \pm\left[\frac{1}{\Delta^{L-2}}-
\frac{1}{\Delta^{L}}\right] ~+~  \mathcal{O}(\Delta^{-L-1})
\label{two_im_pert}
\end{equation}
The two solutions are nearly degenerate with the splitting decreasing exponentially with $L$.
The argument above, i.e. matching of the behaviors of the two sides of~\eref{1p_eqns_im} in the
large $L$ limit, is similar to the argument used originally by Bethe to argue the presence of string
solutions \cite{skorik-95, bethe-31}.

The analytic result for $k_\pm$, Eq.~\eref{two_im_pert}, compares well with the exact numerical
solution of~\eref{1p_eqns_im} at large $\Delta$, as seen in Figure~\ref{fig_1}(b).  The
large-$\Delta$ result \eref{two_im_pert} does not include the vanishing behavior of $k_+$ at
$\Delta_e$, but for $k_-$ it works reasonably down to $\Delta$ values lower than $\Delta_e$.

\subsection{The exceptional point $\Delta_e$}
\label{excep_point}

In the vicinity of $\Delta_e$, we numerically observe $\textrm{Im}(k_+)\sim c_+
(\Delta-\Delta_e)^{1/2}$.  Plugging this ansatz into Eq.~\eref{1p_eqns_im}, 
Taylor expanding both sides in $(\Delta-\Delta_e)$, and collecting the coefficient of
$(\Delta-\Delta_e)^{1/2}$, one obtains 
\begin{equation}
\Delta_e ~=~ \frac{L+1}{L-1}
\label{excep_exact}
\end{equation}
Considering the next non-zero order in $(\Delta-\Delta_e)$, we further obtain
\begin{equation}
c_+=\sqrt{\frac{6(L-1)}{L(L+1)}}
\end{equation}

It turns out that the vanishing real momentum in Fig.~\ref{fig_1} (a) is described by the same
function $\sim c_+(\Delta_e-\Delta)^{1/2}$.  Comparison between the analytic expansions obtained
above and the numerical solutions of the Bethe equations are shown in Fig.~\ref{fig_1} (a) and (b).
From~\eref{excep_exact}, we note that $\Delta_e\to 1$ at $L\to\infty$, signaling that edge-locking
persists in both eigenstates in the complete region $\Delta\in(1,\infty)$.

\begin{figure}[t]
\begin{center}
\includegraphics[width=0.85\textwidth]{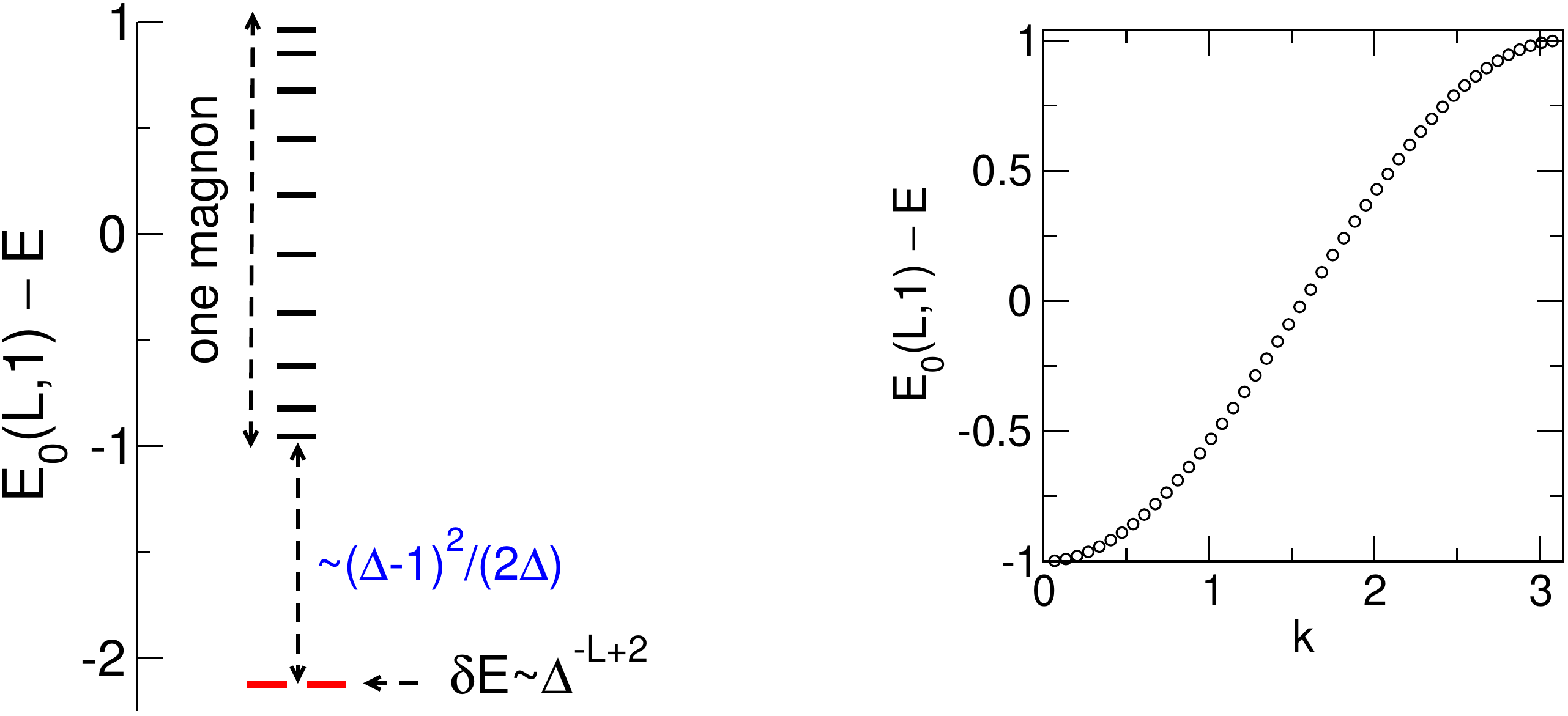}
\end{center}
\caption{ Energy spectrum of the open XXZ chain in the $M=1$ sector at $\Delta=4$, obtained from
  numerical solution of the Bethe equations.  Left: $L=12$.  We invert the energies and add
  $E_0(L,1)\equiv (L-5)\Delta/4$ so that the magnon (delocalized) part of the spectrum appears in
  the $(-1,1)$ range, and the edge-locked doublet appears at the bottom.  The doublet is separated
  from the magnon states by a gap that scales as $\sim(\Delta-1)^2/ (2\Delta)$ at large $\Delta$.
  Right panel:  $E_0(L,1)-E$ versus the Bethe momentum $k$ for the magnon states
  (levels above the gap in the left figure).  Data for a chain with $L=48$ at $\Delta=4$.  }
\label{fig_3}
\end{figure}

\subsection{Signature of edge-locking in the energy spectrum}

We now discuss the spectrum of the open-boundary XXZ chain in the one particle sector. In
Figure~\ref{fig_3} we show $E_0(L,1)-E$ at $\Delta=4$; here $E_0(L,1)\equiv \frac{1}{4}(L-5)\Delta$. Note
that $4>\Delta_e=13/11$.

The half-strings $k_\pm$ correspond to a doublet of quasi degenerate energies at the bottom of the
inverted spectrum and are separated by a gap from higher levels.  Energy levels above the gap
correspond to real solutions of the Bethe equations and exhibit the one-magnon
dispersion~\eref{1p_ener_lambda} (similar to periodic boundary conditions). This is shown in more
detail in the inset for a chain with $L=48$ plotting $E_0(L,1)-E$ versus the Bethe momentum $k$.

The gap scales as $\sim\frac{1}{2}\Delta$ for large $\Delta$ and does not depend on the system size,
as expected for a surface localized state.  More precisely the energy of the edge-locked doublet
(using~\eref{1p_ener_lambda} and~\eref{two_im_pert}) is given as $E_0(L,1)-E\sim
-(1+\Delta^2)/(2\Delta)$, implying that the distance from the bottom of the one magnon band
(obtained at $k\sim 0$) is $\sim (\Delta-1)^2/(2\Delta)$.  The splitting between the two levels
forming the doublet decreases exponentially with the system size (as $\Delta^{-L+2}$).

\subsection{Edge-locked eigenfunctions}

\begin{figure}[t]
\begin{center}
\includegraphics[width=.9\textwidth]{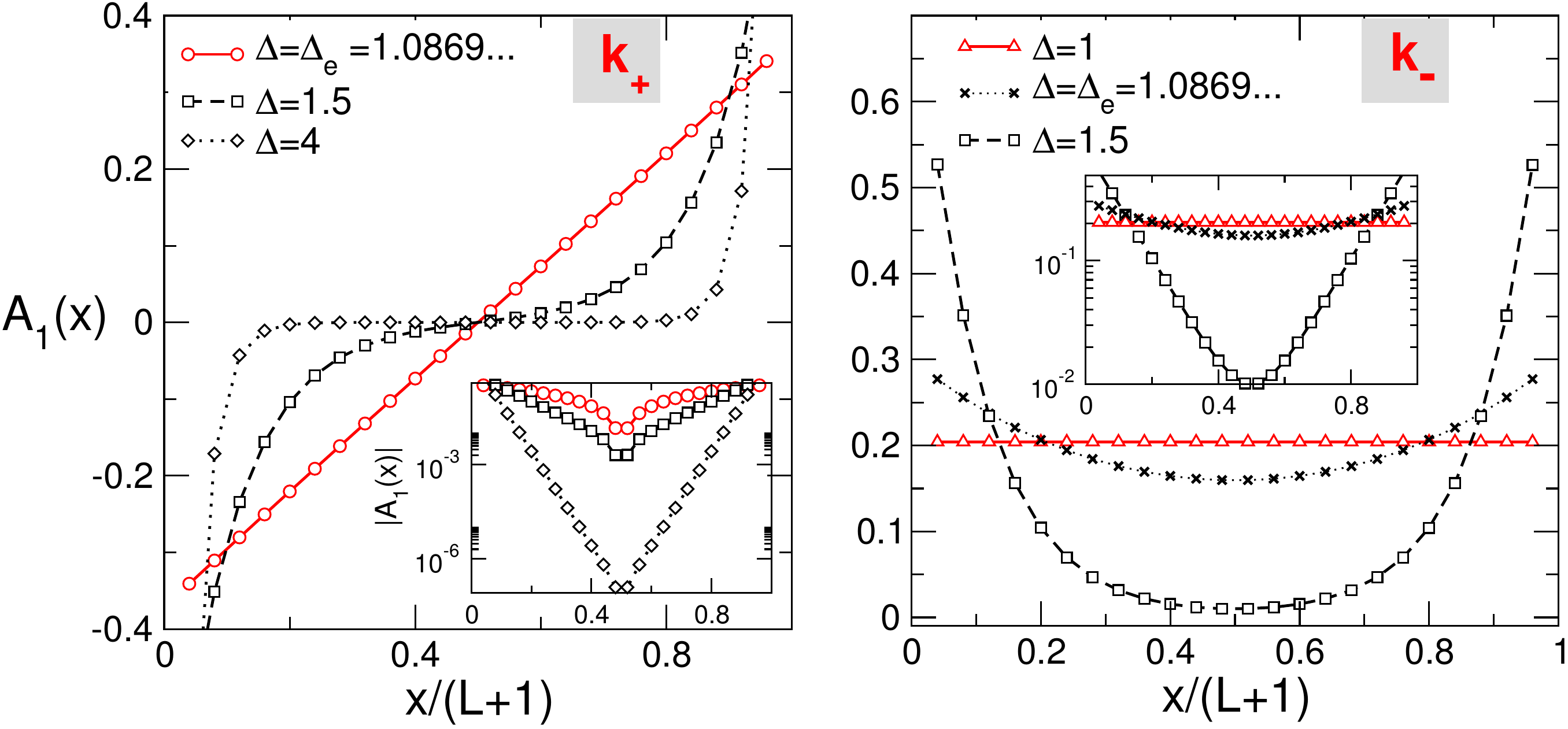}
\end{center}
\caption{ Spatial structure of edge-locked one-particle eigenstates.  Wavefunction amplitude
  $A_1(x)$, obtained from the Bethe equations for chain length $L=24$, are plotted against
  $x/(L+1)$, $x$ being the particle position.  Left and right panels correspond respectively to the
  imaginary solutions $k_+$ and $k_-$.  Insets demonstrate exponential localization by plotting same
  data as in main figures on log-linear scale.  Left ($k_+$) panel: continuous red line is the
  result at the exceptional point $\Delta_{e}=\frac{L+1}{L-1}$, where this imaginary solution
  becomes real.  Right ($k_-$) panel: continuous red line is the wavefunction at the isotropic point
  $\Delta=1$, where this imaginary solution becomes real.  }
\label{fig_4}
\end{figure}

We now highlight the edge-locked nature of the doublet at the bottom of the inverted energy spectrum
in Figure~\ref{fig_3} by analyzing the corresponding eigenfunctions.

These are shown in Figure~\ref{fig_4}, plotting wavefunction components $A_1(x)$ against $x/(L+1)$,
with $x\in[1,L]$ being the position of the particle in the chain.  We consider the eigenfunctions
obtained from the two imaginary solutions $k_+$, $k_-$.

Both eigenfunctions exhibit exponential localization at the edges of the chain at
$\Delta>\Delta_e=25/23\sim1.0869$.  Their form in the large $\Delta$ limit is
$\psi(k_\pm)=(|1\rangle\mp |L\rangle)/\sqrt{2}$, $|1\rangle,|L\rangle$ being the configurations with
the particle localized at the first and last site of the chain. The superposition is a consequence
of the left-right symmetry of the chain, i.e., symmetry under inversion, ${\mathcal I}:i\to L-i+1$.

The eigenfunction obtained from $k_-$ is localized for any $\Delta>1$, and it becomes ``flat'' in
the limit $\Delta\to 1$.  The eigenfunction corresponding to $k_+$ becomes ``linear'' at
$\Delta\to\Delta_e$.  This observation provides an alternative way of determining the value of the
exceptional point $\Delta_e$. Let us consider the ``linear'' superposition
\begin{equation}
 |\psi\rangle=\alpha\sum\limits_{x=1}^L c_x |x\rangle,\qquad c_x\equiv
x-(L+1)/2 , \qquad \alpha\equiv (\sum_{x=1}^L|c_x|^2)^{-1/2} 
\label{ex_eig}
\end{equation}
with $|x\rangle$ denoting the configuration with particle at position $x$.  Now one can require
that~\eref{ex_eig} is an exact eigenstate of the XXZ hamiltonian~\eref{xxz_ham} with eigenvalue
(from Figure~\ref{fig_1} it is $k_+\to0$ at $\Delta\to\Delta_e$)

\begin{equation}
 E=1+E_0(L,1),\qquad E_0(L,1)\equiv (L-5)\Delta_e/4
\label{ener_crit}
\end{equation}

Using the Schr\"odinger equation, one can write 

\begin{equation}
 E\equiv\langle \psi|H|\psi\rangle=|\alpha|^2(\sum_{x=2}^{L-1}c_x
 c_{x+1}+\Delta_e (c_1^2+c_L^2)/2)+\Delta_e (L-5)/4
 \label{eig_der}
\end{equation}

and equating~\eref{ener_crit} and~\eref{eig_der} one obtains  $\Delta_e$ 
as 

\begin{equation}
 \Delta_e=2\frac{\sum_{x=1}^L |c_x|^2-\sum_{x=2}^{L-1}c_xc_{x+1}}
 {(c_1^2+c_L^2)}=(L+1)/(L-1)
\end{equation}

which is the same result obtained in~\ref{excep_point}.

\section{The two particle sector: overview}
\label{two_part}

The rest of the Article, from this section up to Section \ref{sec_2p_spectrum}, details the $M=2$
sector.  In this section, we provide an overview, outlining the different types of two-particle
eigenstates.

\paragraph*{Bethe equations.}

Each eigenstate in the two-particle sector is labeled by two Bethe momenta $k_1$ and $k_2$, whose
possible values are given by two coupled equations:
\begin{equation}
\fl e^{i2(L+1)k_1}\frac{(1-\Delta e^{-ik_1})^2}{(1-\Delta e^{ik_1})^2}=
\frac{(1+e^{i(k_1+k_2)}-2\Delta e^{ik_1})
(1+e^{i(-k_1+k_2)}-2\Delta e^{ik_2})}{(1+e^{i(k_1+k_2)}-2\Delta e^{ik_2})
(1+e^{i(-k_1+k_2)}-2\Delta e^{-ik_1})} 
\label{2p_bethe_eq1}
\end{equation}
\begin{equation}
\fl  e^{i2(L+1)k_2}\frac{(1-\Delta e^{-ik_2})^2}{(1-\Delta e^{ik_2})^2}=
\frac{(1+e^{i(k_1+k_2)}-2\Delta e^{ik_2})(1+e^{i(-k_2+k_1)}-2\Delta 
e^{ik_1})}{(1+e^{i(k_1+k_2)}-2\Delta e^{ik_1})(1+e^{i(-k_2+k_1)}-2\Delta 
e^{-ik_2})}
\label{2p_bethe_eq2}
\end{equation}
The two equations are related by exchange of $k_1$ and $k_2$.
The energy eigenvalue for the generic two-particle eigenstate reads
\begin{equation}
 E=\cos(k_1)+\cos(k_2)+E_0(L,2),\qquad E_0(L,2)\equiv (L-9)\Delta/4
\label{o_energy_2p}
\end{equation}

\paragraph*{Types of solutions.}

The condition that the energy~\eref{o_energy_2p} is real, allows four possibilities for the momentum
pairs $(k_1,k_2)$: 
\begin{eqnarray*}
\textrm{Im}\, (k_1)=\textrm{Im}\,(k_2)=0  \qquad & \textrm{both real} \\
\textrm{Re}\,(k_1)=\textrm{Re}\,(k_2)=0  \qquad & \textrm{both imaginary} \\
\textrm{Im}\,(k_1)=\textrm{Re}\,(k_2)=0  \qquad & \textrm{one real and one imaginary} \\
k_1=k_2^*   \qquad & \textrm{complex conjugate pair (string)} 
\end{eqnarray*}
Note that the trivial solutions $k_1=k_2=0$ and $k_1=k_2=\pi$ have to be discarded and one can
restrict to the region of the complex plane depicted in Figure~\ref{fig_0}.

The physical meanings of the four different types of eigenstates are illustrated via real-space
configurations shown in Table \ref{table_1}.  At large $\Delta$, the edge-localization or mutual
binding is strong.  The eigenstates are then closely represented by the types of configurations
shown in Table \ref{table_1}.  Therefore, at large $\Delta$ we can use combinatorial arguments on
the number of available configurations to count the number of eigenstates of different type.  This
is shown in the second column of Table \ref{table_1}.

\begin{table}[tp]
\begin{center}
\caption{Classification of the $\frac{1}{2}L(L-1)$ two-particle eigenstates at large $\Delta$ and
  large $L$.  Particle configurations (left column) and number of eigenstates (center column) for
  each type of solution of the Bethe equations (right column). $\textrm{Re}$ ($\textrm{Im}$) stands
  for pure real (imaginary) momenta.  Symbols $\fm$ and $\nm$ denote bulk particles and boundary
  particles respectively.  Symbols $\otimes$ denote sites where particles are not allowed.
  Subscripts denote fixed positions.  First row: states that do not exhibit edge-locking.  Second
  row: states with one edge-locked particle and one extended particle. Third row: fully edge-locked
  states.}
\label{table_1}
\vspace{.2cm}
\begin{tabular}{c c c }
\toprule
\T\B {\bf configurations}     & {\bf Number of states}    & {\bf Momenta} 
$\mathbf{\{k_1,k_2\}}$\\

\hline
\T\B$|\otimes_1\cdots\fm
\cdots\otimes\cdots\fm
\cdots\otimes_L\rangle$       & $(L-3)(L-4)/2$            & \{Re, Re\}\\
\T\B$|\otimes_1\cdots\fm\fm
\cdots\otimes_L\rangle$       & $L-3$                     & String\\

\hline
\T$|\nm_1\otimes_2\cdots\fm
\cdots\otimes_L\rangle$       &                           & \\
$\pm$                         & $2(L-3)$                  & \{Re, Im\}\\
\B$|\otimes_1\cdots
\fm\cdots\otimes_{L-1}
\nm_L\rangle$                 &                           &\\

\hline
\T$|\nm_1\nm_2\cdots\rangle
+|\cdots\nm_{L-1}\nm_L
\rangle$                      & 1                         & \\
$|\nm_1\cdots\nm_L\rangle$    & 1                         & \{Im, Im\}\\
\B$|\nm_1\nm_2\cdots\rangle
-|\cdots\nm_{L-1}\nm_L
\rangle$                      & 1                         & \\
\bottomrule

\end{tabular}
\end{center}
\end{table} 

The \{Re,Re\} solutions correspond to eigenstates where the two particles are not locked at the
edges and not bound to each other.  The number of such solutions/eigenstates is therefore the
number of ways of placing the two particles in non-adjacent, non-boundary, sites, hence
$\frac{1}{2}(L-3)(L-4)$.  The number of string (complex conjugate) solutions --- eigenstates where
the two particles are extended but mutually bound --- is the number of ways of placing the two
particles in neighboring non-boundary sites, hence $(L-3)$.  The \{Re,Im\} solutions correspond to
eigenstates where one particle is edge-locked and the other is extended.  The number of such
solutions is the number of ways one can place a particle at one edge and the other in a
non-adjacent, non-edge position, hence $2(L-3)$.  Finally, there are three fully edge-locked
eigenstates given by \{Im,Im\} solutions --- two with a bound pair at the same edge and one with the
two particles at two edges.  Of course, the eigenstates all have definite parity under reflection;
we have therefore used linear combinations of left-edge-locked and right-edge-locked configurations
where appropriate.  For example, of the three fully edge-locked eigenstates, two are symmetric under
${\mathcal I}:i\to L-i+1$, and one is antisymmetric.

\paragraph{Smaller $\Delta$.}

The scenario outlined in Table~\ref{table_1} is true for large $\Delta$.  Since the localization
length scale increases with decreasing $\Delta$, it is perhaps not very surprising that this picture
gets modified at smaller anisotropies.  For each fixed $L$ we find that the number of edge-locked
particles changes as function of $\Delta$.  While in the one particle sector ($M=1$) this
corresponds to a transition from one pure imaginary to real momentum (``delocalization'' of the
edge-locked state), here a richer scenario appears with several types of delocalization transitions
between the different classes of momentum pairs in Table~\ref{table_1}.

For each fixed $L$ two of the three fully edge-locked states 
($\{\textrm{Im}$, $\textrm{Im}\}$ type in Table~\ref{table_1}) decay at 
two distinct exceptional points $\Delta_{e,1}^{(ii)}<\Delta_{e,2}^{(ii)}$. 
As $\Delta$ approaches the exceptional point  one of the two 
imaginary momenta forming the pair vanishes and emerges on the other side (of the exceptional point)
as a real momentum, i.e. the transformation $\{\textrm{Im},\textrm{Im}\}
\to\{\textrm{Re},\textrm{Im}\}$ occurs. The resulting pair $\{\textrm{Re},\textrm{Im}\}$ 
survives at $\Delta>1$. Interestingly, the remaining edge-locked state 
disappears only in the limit $\Delta\to 1$. 

Similar behavior is shown by the type $\{\textrm{Re},\textrm{Im}\}$. At fixed $L$ half of the states
(i.e. $L-3$ states) become magnon-like upon lowering $\Delta$, i.e. one has the transformation
$\{\textrm{Re},\textrm{Im}\}\to\{\textrm{Re},\textrm{Re}\}$.  These transitions occur at $L-3$ (one
for each state) exceptional points $\Delta^{(r,i)}_{e,\ell}$ with $\ell=1,2,\dots,L-3$.  The
remaining $L-3$ edge-locked states decay only in the limit $\Delta\to 1$.  Finally, the string
states are found to be stable in the whole range $1<\Delta<\infty$, i.e., there is no unbinding.

It is worth stressing that for periodic boundary conditions the  
structure of solutions of the Bethe equations outlined in Table~\ref{table_1} 
becomes strikingly simpler. In fact only the first row survives and one has, 
at least in the large $\Delta$ regime, $L$ complex conjugate momenta
(strings) and $L(L-3)/2$ real momentum pairs.

\section{Extended states of two non-bound particles: real momentum pairs}
\label{sec_twoparticle_bothreal}

In this section we focus on the set of real solutions of
Eqs.~\eref{2p_bethe_eq1}\eref{2p_bethe_eq2}, i.e. \{Re,Re\} in Table~\ref{table_1}). To this purpose
it is convenient to rewrite the Bethe equations in logarithmic form. Taking the logarithm on both
sides of~\eref{2p_bethe_eq1}\eref{2p_bethe_eq2} one obtains

\begin{eqnarray}
\fl (L+1)k_1=&\pi J_1 -2 \arctan\frac{\Delta \sin k_1}{1-\Delta 
\cos k_1}-\arctan\frac{\Delta \sin(\frac{k_1+k_2}{2})}
{\cos(\frac{k_1-k_2}{2})-\Delta\cos(\frac{k_1+k_2}{2})}\nonumber\\
& -\arctan\frac{\Delta \sin(\frac{k_1-k_2}{2})}{
\cos(\frac{k_1+k_2}{2})-\Delta\cos(\frac{k_1-k_2}{2})}
\label{loga}\\\nonumber
\fl (L+1)k_2=&\pi J_2 -2 \arctan\frac{\Delta\sin k_2}{1-\Delta 
\cos k_2}-\arctan\frac{\Delta\sin(\frac{k_2+k_1}{2})}
{\cos(\frac{k_1-k_2}{2})-\Delta\cos(\frac{k_2+k_1}{2})}\nonumber\\
&-\arctan\frac{\Delta\sin(\frac{k_2-k_1}{2})}{\cos(\frac{k_1+k_2}{2})
-\Delta\cos(\frac{k_2-k_1}{2})}\label{logb}
\end{eqnarray}

In contrast to the one particle case, we do not redefine the Bethe momenta in terms of the rapidity
variables $\lambda$.  Here $J_1$ and $J_2$ are the Bethe quantum numbers ($J_1,J_2\in[1,L]$).  Since
exchanging $J_1$ and $J_2$ has the effect of swapping $k_1\leftrightarrow k_2$, and since $J_1=J_2$
would imply $k_1=k_2$, one can restrict to $J_1<J_2$.  The counting of the remaining possibilities
gives $L(L-1)/2$ possible real solutions for~\eref{logb}. We anticipate that as for periodic
boundary conditions, however, only some pairs $(J_1,J_2)$ of Bethe numbers give real solutions. Note
also that for periodic boundary conditions one would have the condition $k_1+k_2=2\pi/L(J_1+J_2)$
(conservation of total momentum), reducing the analog of equations~\eref{loga}\eref{logb} to a
single variable equation.

\begin{figure}[t]
\begin{center}
\includegraphics[width=.95\textwidth]{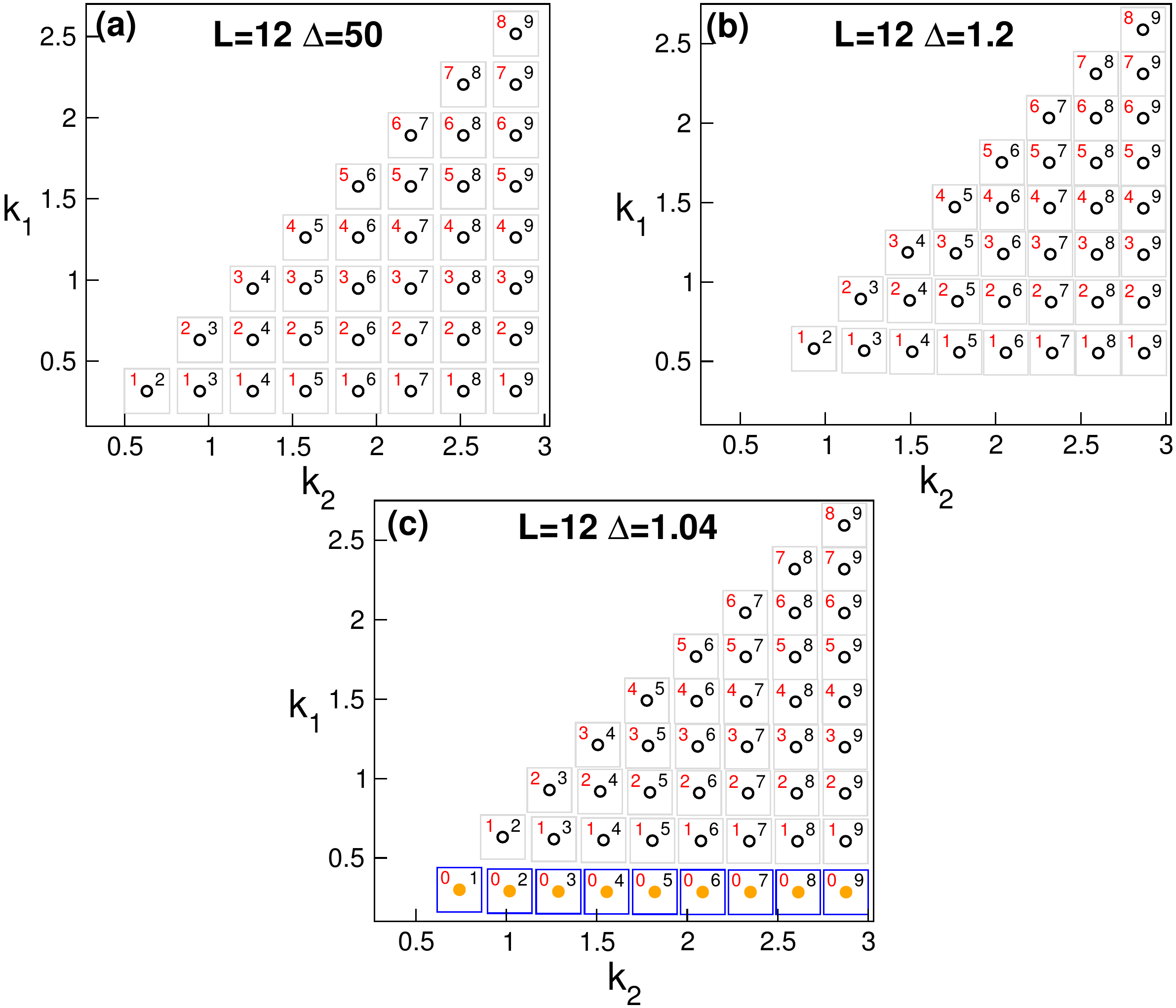}
\end{center}
\caption{ Real solutions ($k_1$,$k_2$) of the Bethe equations for the two-particle ($M=2$) sector;
  $L=12$ chain.  The ($k_1$,$k_2$) values are shown as circles and the accompanying numbers inside
  boxes are the corresponding Bethe quantum numbers $(J_1,J_2)$.  The same Bethe numbers appear with
  the real solutions at $\Delta=50$ and $\Delta=1.2$; panels (a) and (b).  Panel (c) shows a
  $\Delta$ value below the region of exceptional points $\Delta_e^{(rr)}$.  An extra row of real
  momentum pairs with Bethe numbers $J_1=0$, $J_2=1,2,\dots,L-3$ is now present.  }
\label{fig_5}
\end{figure}

We first observe that the first two terms in the r.h.s. of~\eref{loga} \eref{logb} are not
continuous as function of $k_1,k_2$ in the interval of interest, $k_1,k_2\in(0,\pi)$. This can be
avoided by means of the following redefinitions

\begin{equation}
\label{red_1}
\fl\arctan\frac{\Delta \sin k_i}{1-\Delta 
\cos k_i}\to \arctan\frac{\Delta \sin k_i}{1-\Delta 
\cos k_i}-\pi\theta(k_i-\arccos(1/\Delta))
\end{equation}

\begin{eqnarray}
\label{red_2}
\fl \arctan\frac{\Delta\sin(\frac{k_i+k_j}{2})}
{\cos(\frac{k_i-k_j}{2})-\Delta\cos(\frac{k_i+k_j}{2})}
\to & \arctan\frac{\Delta\sin(\frac{k_i+k_j}{2})}
{\cos(\frac{k_i-k_j}{2})-\Delta\cos(\frac{k_i+k_j}{2})}-\\
\nonumber & \pi\theta\left( k_j-2\arccos\frac{(1+\Delta)\sin
\frac{k_i}{2}}{\sqrt{1+\Delta^2-2\Delta\cos k_i}}\right)
\end{eqnarray}

Note the presence of the Heavside step functions $\theta(x)\equiv(1+
\textrm{sign}(x))/2$ in~\eref{red_1}\eref{red_2}, which contribute with 
a $\pm\pi$ phase shift (depending on the interplay between the Bethe 
momenta and $\Delta$) in the Bethe equations.  This amounts to a redefinition  
of the Bethe numbers $J_1,J_2$, and it  is simple to understand in the Ising 
limit $\Delta\to\infty$.  One has then 
\begin{equation}
\label{red_1_a}
\arctan\frac{\Delta \sin k_i}{1-\Delta 
\cos k_i}\to-k_i+\pi\theta(2k_i-\pi)
\end{equation}
\begin{equation}
\label{red_2_a}
\arctan\frac{\Delta \sin(\frac{k_i+k_j}{2})}
{\cos(\frac{k_i-k_j}{2})-\Delta\cos(\frac{k_i+k_j}{2})}
\to-\frac{k_i+k_j}{2}+\pi\theta(k_j+k_i-\pi)
\end{equation}
implying that, given a solution $k_i$ of the Bethe equations, 
the corresponding Bethe number $J_i$ is shifted by $1$ if 
$k_i>\pi/2$ (cf.~\eref{red_1_a}). Another additional shift is obtained 
if $k_i+k_j>\pi$ (cf.~\eref{red_2_a}). 

The Bethe momenta $k_1,k_2$, obtained from numerical solutions of \eref{loga}, \eref{logb}, with the
redefinitions~\eref{red_1}, \eref{red_2}, are shown in Figure~\ref{fig_5} for a $L=12$ chain, for
three $\Delta$ values.  For each pair $(k_1,k_2)$ the corresponding Bethe quantum numbers $(J_1,J_2)$
are also reported.

At $\Delta=50$ the Bethe momenta appear to be ``quantized'' in units of $\pi/(L-2)$ forming a
triangular structure in the plane $k_2,k_1$.  Moreover ``bands'' of quasi degenerate momenta $k_1$
and $k_2$ are present, respectively ``rows'' and ``columns'' of solutions in the Figure. Solutions
within the same row (column) have the same quantum number $J_1$ ($J_2$).  A similar structure
persists at much lower values of $\Delta$ as seen for $\Delta=1.2$ in Figure~\ref{fig_5}(b).  The
same triangular structure is observed apart from deviations at small $k_1$, $k_2$.

The simple structure of the Bethe numbers observed in Figure~\ref{fig_5} depends crucially
on~\eref{red_1}, \eref{red_2}.  A striking consequence of using~\eref{red_1}, \eref{red_2} is that
{\it all} the pairs of Bethe numbers $1\le J_1<J_2\le L-3$ correspond to real solutions of the Bethe
equations. Also the set of Bethe numbers does not depend on the anisotropy $\Delta$ (the same
integer pairs $(J_1,J_2)$ give the Bethe momenta of both Figure~\ref{fig_5} (a) and (b)).  With a
different redefinition (different choice of branches of the $\arctan$ function), the set of $J_1$,
$J_2$ values giving real solutions can be different.

Finally, Figure~\ref{fig_5}(c) plots the solutions of the Bethe equations at $\Delta=1.04$.  Now, we
find that $L-3$ extra real momentum pairs appear as a new row at the bottom of the triangle; these
correspond to Bethe numbers $J_1=0$ and $J_2=1,2,\dots, L-3$. At large $\Delta$ these extra real
solutions undergo the transformation $\{\textrm{Re}, \textrm{Re}\}\to\{\textrm{Re},\textrm{Im}\}$
and disappear.  This transformation is discussed in the next subsection.

\subsection{\{Re,Re\}$\to$\{Re,Im\} transformations: edge-locking of a single particle at exceptional points}

\begin{figure}[t]
\begin{center}
\includegraphics[width=.9\textwidth]{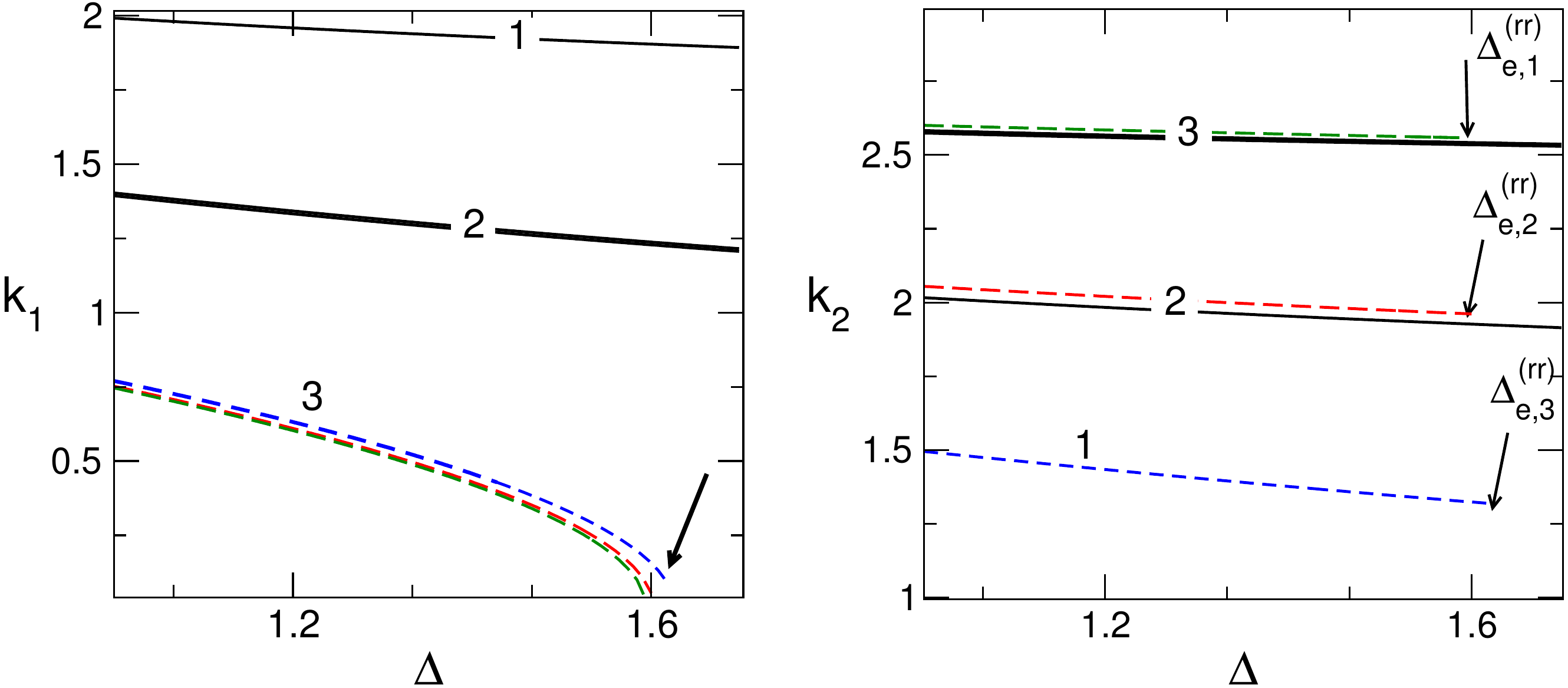}
\end{center}
\caption{ Real solutions ($k_1$.$k_2$) for a small ($L=6$) chain as function of anisotropy
  $\Delta$, obtained from numerical solution of the Bethe equations.  The momenta are organized in
  $L-3=3$ ``bands'' at small at $\Delta$ and $L-4=2$ bands at large $\Delta$.  Accompanying integers
  are the numbers of solutions within each band.  The bands of $k_1$ ($k_2$) correspond to rows
  (columns) in Figure~\ref{fig_5}.  The arrows mark the region of exceptional points
  $\Delta^{(rr)}_{e,\ell}\,\,\ell=1,2,3$.  At these points, the $L-3$ solutions corresponding to the
  lowest $k_1$ band at small $\Delta$ disappear.  These correspond to one vanishing value in each
  $k_2$ band, and to the lowest row in Figure~\ref{fig_5}(c).  We use dashed lines for the solutions
  which vanish.  At larger $\Delta$ these solutions will reappear as \{Re,Im\} solutions.  }
\label{fig_7}
\end{figure}

At $\Delta$ values not much higher than $\Delta=1$, the number of real solutions changes, similarly
to what was observed in the $M=1$ sector. Precisely, while $(L-3)(L-2)/2$ real pairs $(k_1,k_2)$ are
present at small $\Delta$, $L-3$ among them undergo the transformation $\{\textrm{Re},\textrm{Re}\}
\to\{\textrm{Re},\textrm{Im}\}$ at the exceptional points
$\Delta^{(rr)}_{e,1}<\Delta^{(rr)}_{e,2}<\dots< \Delta^{(rr)}_{e,L-3}$.  The superscript stresses
that these are the exceptional points for the \{Re,Re\} type of solutions.  In physical terms the
transformation $\{\textrm{Re}, \textrm{Re}\}\to\{\textrm{Re},\textrm{Im}\}$ can be interpreted as
one magnon with vanishing real momentum becoming edge-locked in the limit
$\Delta\to(\Delta^{(rr)}_{e,\ell})^-$. The remaining $(L-3)(L-4)/2$ real pairs survive in the large
$\Delta$ limit (Table~\ref{table_1}).

This scenario is highlighted in Figure~\ref{fig_7} showing the  
momentum pairs $(k_1,k_2)$ (solutions of the Bethe equations) as a 
function of the anisotropy $1\le\Delta<2$  for a chain with $L=6$. 
While $(L-2)(L-3)/2=6$ solutions are present at $\Delta\lesssim 1.6$, 
only $(L-3)(L-4)/2=3$ survive in the large $\Delta$ limit. In particular  
the three solutions $k_1$ forming the lowest ``band'' (cf. 
Figure~\ref{fig_7} (left)) are vanishing in the 
region $\Delta\sim 1.6$ (they are exactly zero at the three exceptional 
points $\Delta^{(rr)}_{e,1},\Delta^{(rr)}_{e,2},\Delta^{(rr)}_{e,3}$). 
These momenta would correspond to the bottom row of solutions in 
Figure~\ref{fig_5} (c). Note that for each vanishing $k_1$ the 
corresponding $k_2$ (right panel in Figure~\ref{fig_7}) is finite at the 
exceptional point.

\begin{figure}[t]
\begin{center}
\includegraphics[width=.9\textwidth]{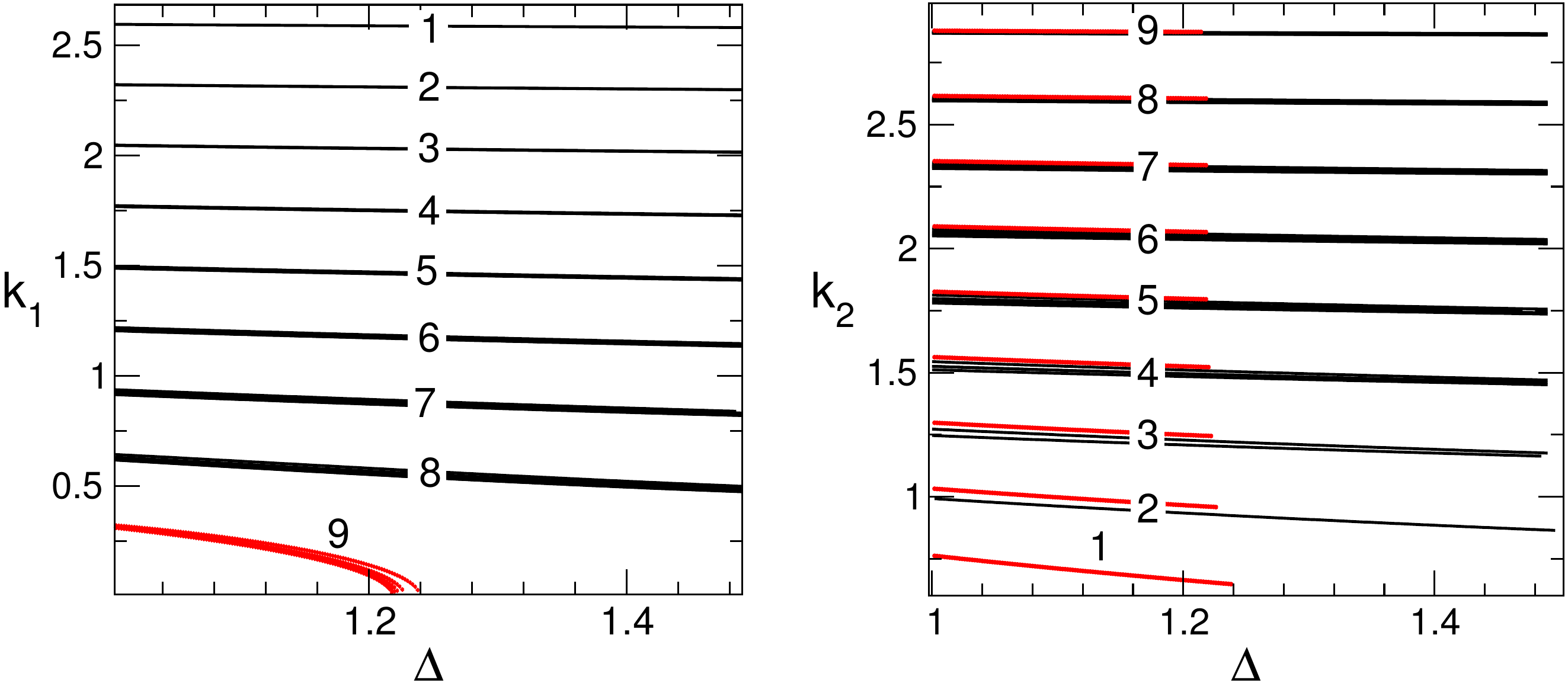}
\end{center}
\caption{ Same as in Figure \ref{fig_7}, now shown for a larger chain ($L=12$).  There are
  $\frac{1}{2}(L-3)(L-4)$ solutions of \{Re,Re\} type at $\Delta$ values above the exceptional
  points (consistent with Table \ref{table_1}) and $\frac{1}{2}(L-2)(L-3)$ such solutions at small
  $\Delta$.  }
\label{fig_8}
\end{figure}

The same scenario outlined so far is  observed at larger $L$. 
In Figure~\ref{fig_8} we show the real Bethe momentum pairs 
$(k_1,k_2)$ for a chain with $L=12$. The same qualitative result 
as in Figure~\ref{fig_7} is found. Also, one has 
$\Delta_{e,1}^{(rr)}<\Delta_{e,2}^{(rr)}<\dots<\Delta_{e,L-3}^{(rr)}
\approx 1.2$, i.e. the exceptional points are nearer to the isotropic 
point $\Delta=1$ (compared to $L=6$), suggesting that $\Delta^{(rr)}_{e,\ell}\to 
1\,\forall\,\ell$ in the limit $L\to\infty$ (as proven in section~\ref{one_p} for 
the single particle sector).

\subsection{The exceptional points}

As in the one particle case (Section~\ref{one_p}), the positions of the exceptional points and the
behavior of the Bethe momenta in their vicinity can be characterized analytically.  We verified that
the momentum pairs $(k_1,k_2)$ disappearing at the exceptional points exhibit the behavior

\begin{equation}
\label{real_fate}
k_1=(\Delta-\Delta_e)^{1/2}+{\mathcal O}(\Delta-\Delta_e)\qquad k_2=\bar 
k+{\mathcal O}(\Delta-\Delta_e)
\end{equation}

as $\Delta\to\Delta_e$.  Here $\Delta_e$ denotes a generic exceptional point
$\Delta^{(rr)}_{e,\ell}$ ($\ell=1,2,\dots,L-3$).  Note that one has from~\eref{real_fate} that $k_1$
is vanishing while $k_2$ assumes the finite value $\bar k$ at $\Delta_e$.  Moreover,
\eref{real_fate} holds on both sides of the exceptional point).  The square root behavior as
$(\Delta-\Delta_e)^{1/2}$ reflects the transformation from real to pure imaginary of the momentum
$k_1$, i.e., the edge-locking transformation.

After substituting~\eref{real_fate} in the Bethe equations, one obtains that 
$\bar k,\Delta_{e}$ are determined by a set of coupled  equations 
as 
\begin{eqnarray}
\label{ex_re_re}
\cos\bar k=\frac{-1+\Delta_e+2\Delta_e^2-4\Delta_e^3+L(\Delta_e-1)
(1+2(\Delta_e-1)\Delta_e)}{1+L+\Delta_e-3L\Delta_e+2(L-2)\Delta_e^2}
\\
\label{ex_re_re_2}
 e^{2i\bar k L}\frac{(e^{i\bar k}-\Delta_e)^2}
{(e^{-i\bar k}-\Delta_e)^2}-\frac{(1+e^{i\bar k}
-2e^{i\bar k}\Delta_e)^2}{(1+e^{-i\bar k}-2\Delta_e e^{-i\bar k})^2}=0
\end{eqnarray}

\subsection{Expansion of the real Bethe momenta in the Ising limit ($\Delta\gg 1$)}

\begin{figure}[t]
\begin{center}
\includegraphics[width=.8\textwidth]{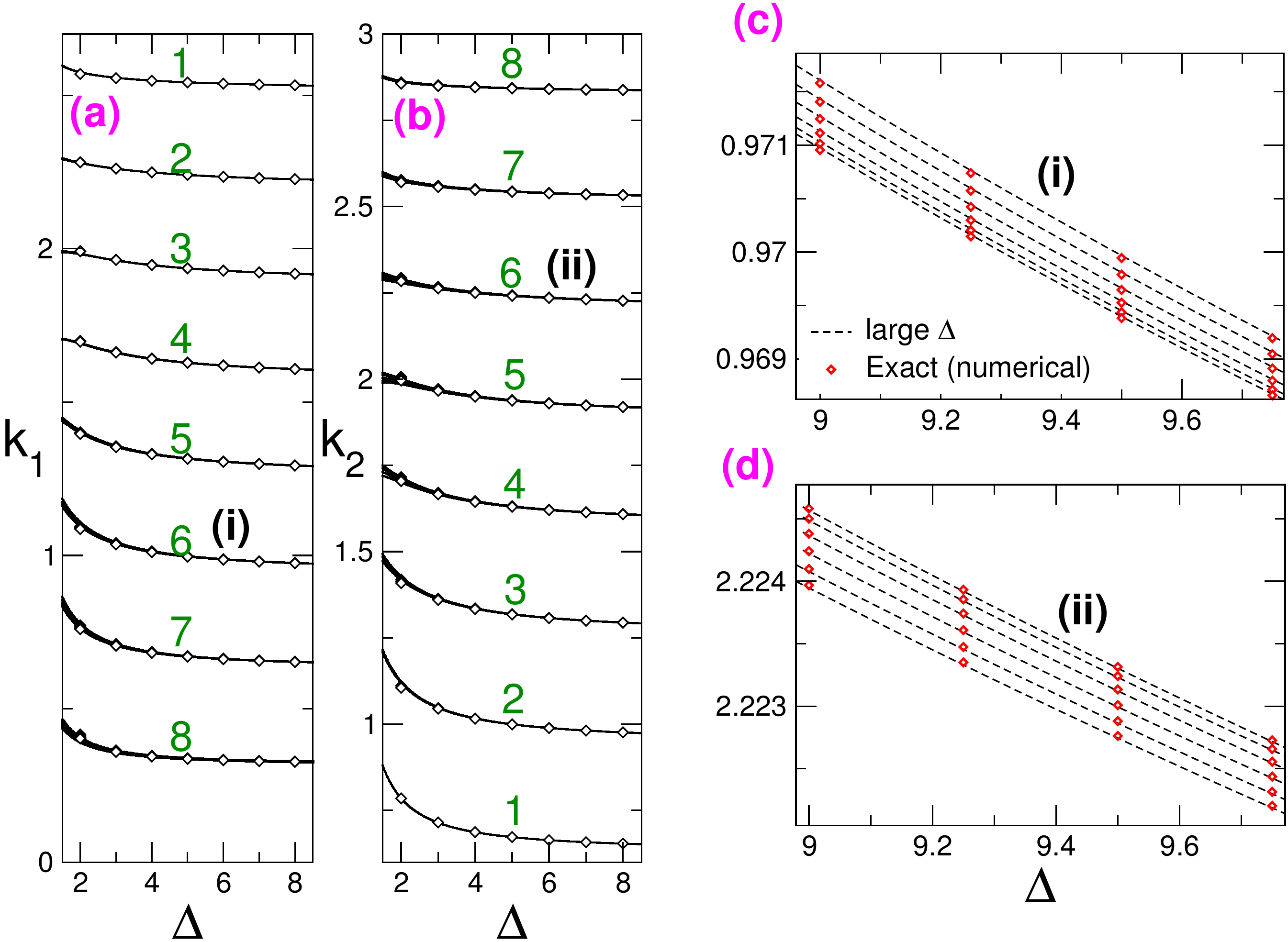}
\end{center}
\caption{ Solutions of \{Re,Re\} type: large $\Delta$ behavior of the real momentum pairs
  $(k_1,k_2)$. Data for $L=12$.  Symbols are numerical solutions of the Bethe equations, and lines
  are the analytic expansion in the large $\Delta$ regime. (\textbf{a},\textbf{b}) The number of
  momenta in each band is shown with the accompanying numbers.  (\textbf{c},\textbf{d}) Zoomed-in
  comparison between exact numerical solutions of the Bethe equations (symbols) and the large
  $\Delta$ result (dashed lines) for the momenta in the $k_1$ band marked as (\textbf{i}) in panel
  (\textbf{a}) and for the momenta in the $k_1$ band marked as (\textbf{ii}) in panel (\textbf{b}).
}
\label{fig_6}
\end{figure}

We now discuss the \{Re,Re\} solutions in the large $\Delta$ regime. 

Guided by the observation that at $\Delta\gg 1$ the Bethe momenta are ``quantized'' in units of
$\pi/(L-2)$ (Figure~\ref{fig_5}(a)), and assuming analytic behavior at finite $\Delta$, one can
expand as

\begin{equation}
\label{large_an}
k_s=k_s^0(J_s)+\sum\limits_{m=1}^\infty a_{s,m}(J_s)\Delta^{-m}, \qquad s=1,2
\label{large_D_exp}
\end{equation}

with $k^0_s$ the zeroth order Bethe momentum given as $k^0_s\equiv
\pi/(L-2)J_s$,  $J_s=1,2,\dots, L-3$ being the Bethe quantum numbers 
(cf. Figure~\ref{fig_5} (a)). The parameters  $a_{s,m}$ are determined by 
substituting~\eref{large_an} in the Bethe equations and equating the 
coefficients of the same powers of $1/\Delta$. One can readily obtain the 
large $\Delta$ expansion of the Bethe momenta up to ${\mathcal O}(\Delta^{-4})$ 
as

\begin{eqnarray}
\label{pert_real}
\nonumber\fl k_1=k^0_1 +\frac{3}{\Delta(L-2)}\sin k^0_1+\frac{1}
{\Delta^2}\frac{((8+5L)\cos k_1^0+(L-2)\cos k_2^0 )\sin k_1^0}{2(L-2)^2}
+\\\fl\frac{1}{4\Delta^3(L-2)^3}\Big[\frac{1}{2}\Big(15-21L+2(L-2)(L+1)
\cos 2k_2^0 \Big)\sin k_1^0 +\\\nonumber\frac{3}{2}(L+1)(5+2L)\sin3 k_1^0  
+(L-2)(L+4)\cos k_2^0\sin 2k_1^0\Big]+{\mathcal O}(\Delta^{-4})
\end{eqnarray}

The expression for $k_2$ is obtained exchanging $1\leftrightarrow 2$ 
in~\eref{pert_real}. The expansion~\eref{pert_real} confirms that  
at the lowest order in $1/\Delta$ the two particles (magnons) are non 
interacting, whereas  interactions start contributing with terms 
$\sim 1/(L\Delta)$.

The large $\Delta$ expansion~\eref{pert_real} is checked in Figure~\ref{fig_6} against exact
numerical solutions of the Bethe equations, for a $L=12$ chain.  For $\Delta\gtrsim2$,
Eq.\ \eref{pert_real} reproduces not only the overall behavior of each band, panels
(\textbf{a},\textbf{b}), but also the fine structure within the bands, panels
(\textbf{c},\textbf{d}).

\section{Fully edge-locked states: pure imaginary momentum pairs}
\label{pure_im}

In this section we focus on the pure imaginary momentum pairs 
$(k_1\equiv i\kappa_1,k_2\equiv i\kappa_2)$ (type $\{\textrm{Im},
\textrm{Im}\}$, three states in the last row  of Table~\ref{table_1}). 
These correspond to eigenstates of~\eref{xxz_ham} with both particles 
locked at the edges of the chain. 

\begin{figure}[t]
\begin{center}
\includegraphics[width=.7\textwidth]{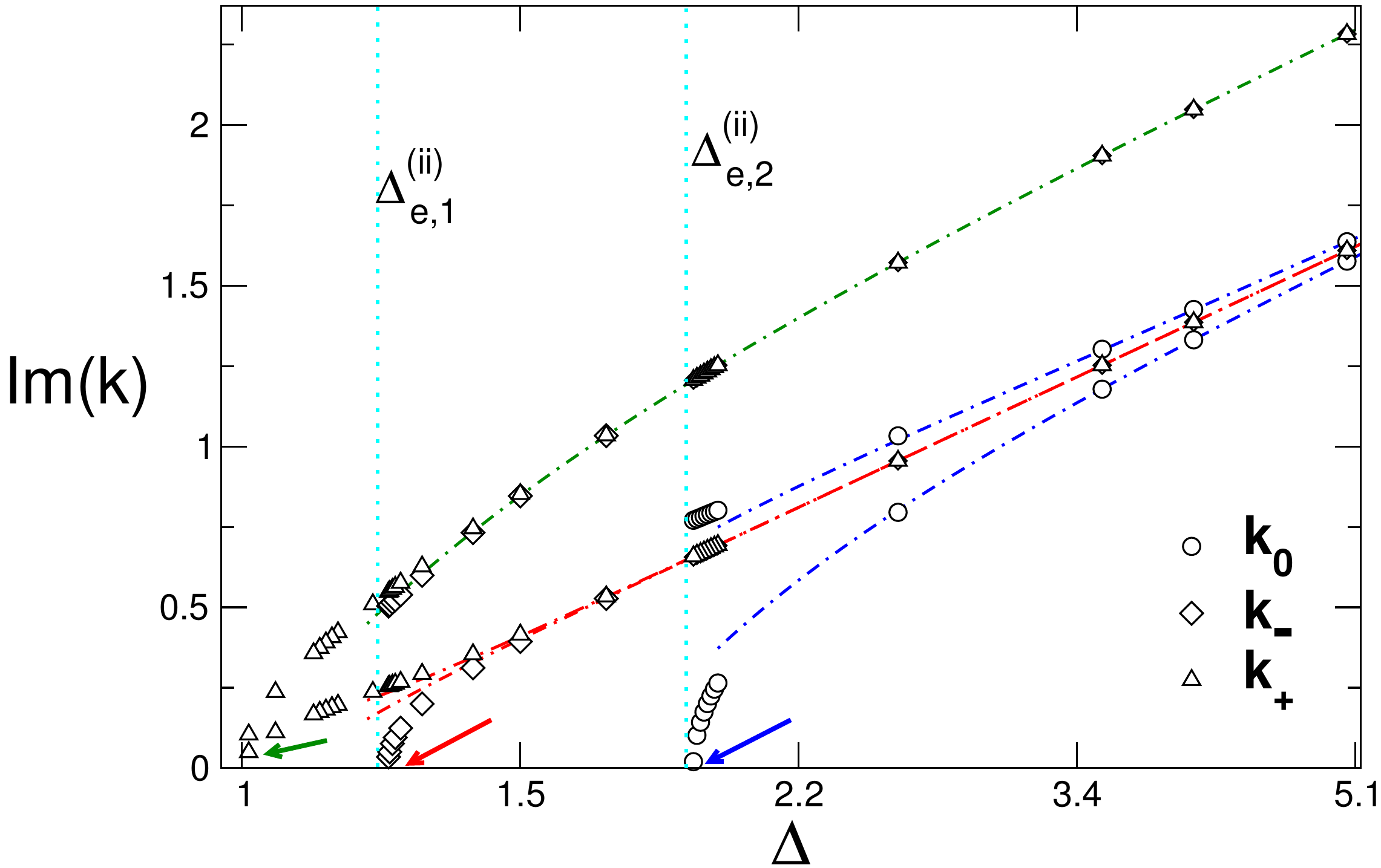}
\end{center}
\caption{ The three solutions of \{Im,Im\} type, denoted ${\mathbf k}_{\pm}$, ${\mathbf
    k}_0$, for an $L=6$ chain.  Momenta forming a pair are denoted with the same symbols.  At large
  $\Delta$ the pairs ${\mathbf k}_\pm$ are ``degenerate''. At the exceptional point
  $\Delta=\Delta_{e,2}^{(ii)}$ ($\Delta_{e,1}^{(ii)}$) one of the two momenta forming the pair
  ${\mathbf k}_0$ (${\mathbf k}_-$) vanishes and the \{Im,Im\} solution disappears.  In the limit
  $\Delta\to 1$ (isotropic point) both the components of ${\mathbf k}_0$ are vanishing, i.e. no pure
  imaginary pairs remain at the isotropic point. Dashed-dotted lines are the analytic expansions at
  large $\Delta$, Eqs.\ \eref{n_deg_theo},\eref{n_deg_theo_6},\eref{large_L_deg}.  }
\label{fig_9}
\end{figure}

Figure~\ref{fig_9} plots all the imaginary momentum pairs (their imaginary parts 
$\kappa_1,\kappa_2$) as function of the anisotropy $\Delta$ for a chain 
with $L=6$ sites. Data are numerical solutions of  the Bethe equations. 
The two components of a given momentum pair  are shown with the same symbols.

Clearly in the large $\Delta$ ($\Delta\gtrsim 2.1$) there are three pure 
imaginary momentum pairs (cf. Table~\ref{table_1}). Two of them 
(that we denote as ${\mathbf k}_\pm$, respectively rhombi and triangles in the Figure) 
are ``degenerate'' at $\Delta\to\infty$, meaning that ${\mathbf k}_+\equiv
(i\kappa_{+,1},i\kappa_{+,2})\to {\mathbf k}_-\equiv(i\kappa_{-,1},i\kappa_{-,2})$.  
On the other hand the ``isolated'' one (${\mathbf k}_0$, circles in the Figure) 
is a pair of two quasi-degenerate momenta in the limit $\Delta\to\infty$. 

At lower $\Delta$ one has two exceptional points $\Delta_{e,1}^{(ii)}, \Delta^{(ii)}_{e,2}$ at which
one component of a pure imaginary pair vanishes. Precisely, this occurs at
$\Delta_{e,2}^{(ii)}\approx 1.9$ for ${\mathbf k}_0$ and at $\Delta_{e,1}^{(ii)} \approx 1.2$ for
${\mathbf k}_-$. Note that ${\mathbf k}_+$ survives up to the isotropic point ($\Delta=1$) where
both its components are vanishing (i.e. $\Delta=1$ is also an exceptional point). The vanishing
momenta at $\Delta\gtrsim\Delta^{(ii)}_{e,\ell}$ emerge on the other side of the exceptional point
(at $\Delta\lesssim\Delta^{(ii)}_{e,\ell}$) as real momenta, i.e.  the transformation
$\{\textrm{Im},\textrm{Im}\}\to\{\textrm{Re}, \textrm{Im}\}$ occurs. This reflects the
delocalization of one of the two edge-locked particles, which becomes extended (magnon-like).

%

Both the behaviors around the exceptional points and in the large $\Delta$ region can be understood
analytically.  It is convenient to parametrize the imaginary Bethe momenta as $z_i\equiv
e^{\kappa_i}$. Now the Bethe equation for $z_1$ reads
\begin{eqnarray}
\label{imag_sim_be}
\fl (z_1-\Delta)^2(z_1+z_2-2\Delta)(1+z_1 z_2-2z_2\Delta)+\\\nonumber
\qquad\qquad (1-z_1\Delta)^2(2\Delta z_2z_1-z_1-z_2)(1+z_1z_2-2\Delta z_1)
z_1^{-2L}=0
\end{eqnarray}
The equation for $z_2$ is obtained from~\eref{imag_sim_be} by exchanging $z_1\leftrightarrow z_2$.
Due to the restriction $\kappa_i>0$ (Section~\ref{sec_1_BA}), one has $z_i>1$, implying that for
large $L$ the second term in~\eref{imag_sim_be} is vanishing exponentially.  From the first term
in~\eref{imag_sim_be} one then has $z_1=z_2=\Delta$ and $z_1=2z_2=2\Delta$ as possible
solutions. The former gives the solution ${\mathbf k}_0\sim (i\log\Delta,i\log\Delta)$, whereas the
latter corresponds to the quasi-degenerate pairs ${\mathbf k}_+\sim {\mathbf k}_-\sim (i\log\Delta+
i\log2,i\log\Delta)$.  Expansions valid up to higher orders in $\Delta^{-L}\ll 1$ are given in the
next subsection.

\subsection{The imaginary momentum pairs at large $\Delta$}

In this section we investigate the fine structure of the pure imaginary momentum pairs
$\mathbf{k}_{\pm}$, $\mathbf{k}_0$ at large $\Delta$.  The small parameter for the expansion is
$\Delta^{-L}$, so that this can also be interpreted as a large $L$ expansion.

We start with the solution $\mathbf{k}_0$ (Figure~\ref{fig_9}).  The idea is to expand the two
members of the imaginary pair as
\begin{equation}
\label{k0_exp}
z^{(0)}_i=\Delta+\sum\limits_{m=0}^\infty\frac{b_{i,m}}{\Delta^{L/2-2+m}},\qquad 
i=1,2
\end{equation}

where the superscript in $z_i^{(0)}$ is to stress that we are focusing on the pair 
${\mathbf k}_0$. The coefficients $b_{i,m}$  are determined by substituting~\eref{k0_exp} 
in the Bethe equations~\eref{imag_sim_be} and solving the linear system obtained 
equating the coefficients of the same powers in $1/\Delta$. After a lengthy 
algebra the first two non trivial orders of $z^{(0)}_1,z^{(0)}_2$ are obtained as 

\begin{eqnarray}
\label{n_deg_theo}
\fl z^{(0)}_{1}=\Delta+\frac{\Delta^2}{(L-3)^{\frac{1}{4}}\Delta^{L/2}}
-\frac{3(L-4)}{[3+(-1)^{L/2-2}]2^{[1-(-1)^{L/2}]/2}(L-3)^{\frac{5}{4}}
\Delta^{L/2}}+o(\Delta^{-L/2})\\
\fl \nonumber z^{(0)}_{2}=\Delta-\frac{\Delta^2}{(L-3)^{\frac{1}{4}}
\Delta^{L/2}}+\frac{3(L-4)}{[3+(-1)^{L/2-2}]2^{[1-(-1)^{L/2}]/2}(L-3
)^{\frac{5}{4}}\Delta^{L/2}}+o(\Delta^{-L/2})
\end{eqnarray}

where $o(\Delta^{-L/2})$ denotes higher order corrections. One should 
stress that the expansion~\eref{n_deg_theo} does not hold if the size of the 
chain is too small. For example, we find that for $L=6$ one has, instead 
of~\eref{n_deg_theo}, the expansion 

\begin{eqnarray}
\label{n_deg_theo_6}
 z^{(0)}_{1}=\Delta+\frac{1}{3^{1/4}\Delta}
-\frac{6+3^{1/4}}{2\cdot 3^{1/2}\Delta^3}+o(\Delta^{-3})\\
 \nonumber z^{(0)}_{2}=\Delta-\frac{1}{3^{1/4}\Delta}
+\left(3^{1/2}-2\cdot 2^{1/2}+\frac{1}{2\cdot 3^{1/4}}\right)
\frac{1}{\Delta^3}+o(\Delta^{-3})
\end{eqnarray}

While the first two orders in~\eref{n_deg_theo_6} for both $z_1^{(0)}$ 
and $z_2^{(0)}$  are correctly reproduced by the general 
result~\eref{n_deg_theo}, this is not 
the case for the last order $\sim\Delta^{-3}$. 

A similar expansion can be carried out for the case of the two almost 
``degenerate'' (at $\Delta\to\infty$) pairs ${\mathbf k}_\pm$ 
(cf. Figure~\ref{fig_9}). The first few orders for the corresponding 
$z^{(\pm)}_\ell$ are given as 

\begin{eqnarray}
\label{large_L_deg}
\fl z^{(+)}_{1}=\Delta+\frac{1}{2^{L-3}\Delta^{2L-5}}+o(\Delta^{-2L+5}) & 
\qquad  z_2^{(+)}=2\Delta-\frac{1}{\Delta}-\frac{1}{(2\Delta)^{2L-5}}+o
(\Delta^{-2L+5})\\\nonumber
\fl z^{(-)}_1=\Delta+\frac{1}{2^{L-3}\Delta^{2L-5}}+ o(\Delta^{-2L+5}) &  
\qquad z_2^{(-)}=2\Delta-\frac{1}{\Delta}+\frac{1}{(2\Delta)^{2L-5}}+o
(\Delta^{-2L+5})
\end{eqnarray}

Note that the higher order corrections decay faster (as $o(\Delta^{-2L+5})$) 
than in the case of ${\mathbf k}_0$ (as $o(\Delta^{-L/2})$).

The validity of the expansions~\eref{n_deg_theo},\eref{n_deg_theo_6},\eref{large_L_deg} for all
pairs ${\mathbf k}_0,{\mathbf k}_\pm$ is checked against exact results obtained by solving the Bethe
equations numerically in Figure~\ref{fig_9}.  It is remarkable that the agreement between the exact
data and the expansions is good even in the region $\Delta\approx 1$. Deviations are only visible
near the exceptional points, where Eqs.\ \eref{n_deg_theo},\eref{n_deg_theo_6},\eref{large_L_deg}
are inadequate.

\subsection{Expansion of the imaginary pairs near the exceptional points}

\begin{figure}[t]
\begin{center}
\includegraphics[width=.9\textwidth]{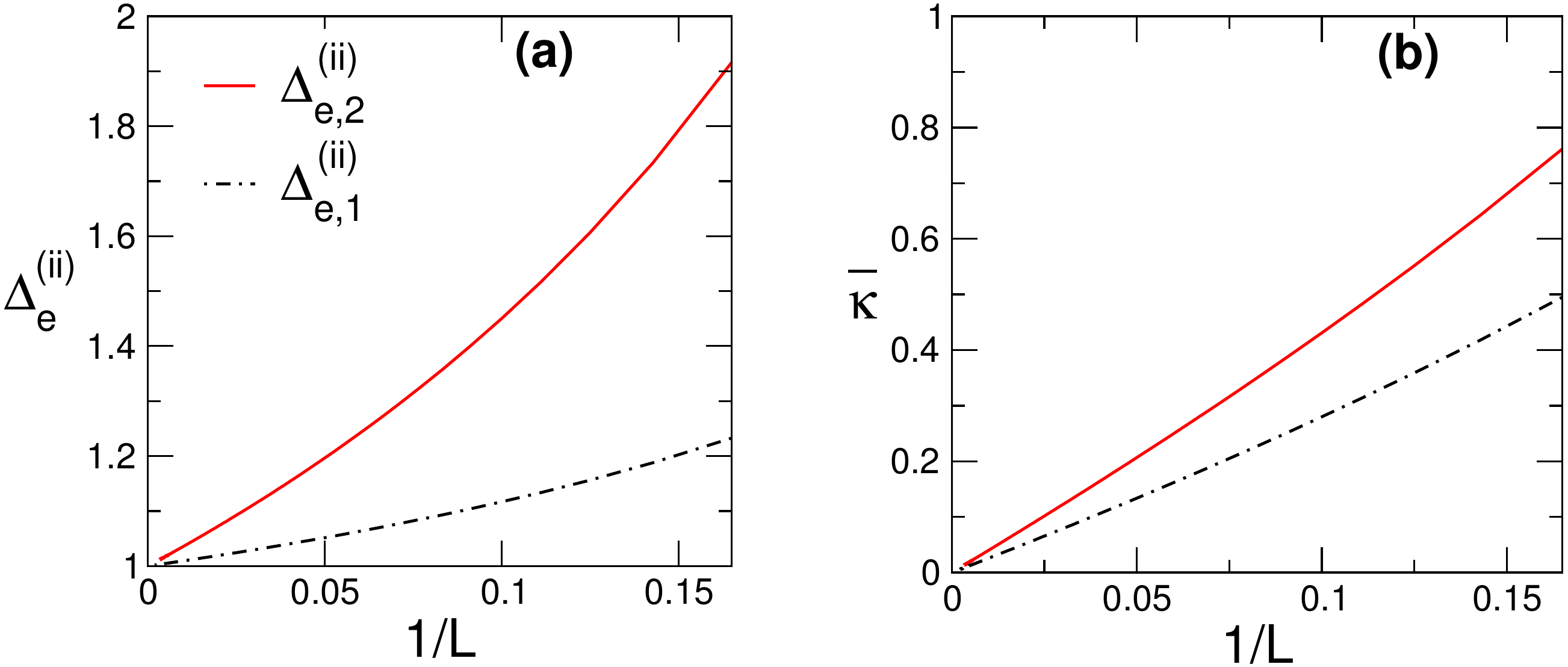}
\end{center}
\caption{ (\textbf{a}) Exceptional points $\Delta_{e,1}^{(ii)}$ and $\Delta^{(ii)}_{e,2}$ as
  function of inverse chain length $1/L$.  Both points coalesce at $\Delta=1$ at large $L$.
  (\textbf{b}) The value $\bar\kappa$ of the non vanishing imaginary momentum at the exceptional
  point.  For both cases $\bar\kappa\to 0$ at large $L$.  }
\label{fig_10}
\end{figure}

The two exceptional points $\Delta_{e,\ell}^{(ii)}$ at which the pure 
imaginary solutions of the Bethe equations for two particles disappear 
are obtained by imposing that the imaginary momentum pair is of the 
form 
\begin{equation}
\label{ansatz_ii}
k_1=i(\Delta-\Delta_e)^{1/2}+{\mathcal O}(\Delta-\Delta_e),  
\qquad k_2=i\bar\kappa+{\mathcal O}(\Delta-\Delta_e)
\end{equation}
with $\Delta_e$ and $\bar\kappa$ respectively the exceptional point and the imaginary value of $k_2$
at $\Delta_{e}$.  Eq.~\eref{ansatz_ii} is valid on both sides of the exceptional point.  It encodes
the transformation \{Im,Im\} $\to$ \{Re,Im\}, i.e., the fact that  $k_1$ is imaginary for $\Delta>\Delta_e$
and real for $\Delta<\Delta_e$.  The two parameters $\Delta_e$ and $\bar\kappa$ are determined by
solving the coupled equations
\begin{eqnarray}
\label{ex_im_im}
\cosh\bar \kappa=\frac{-1+\Delta_e+2\Delta_e^2-4\Delta_e^3+L(\Delta_e-1)
(1+2(\Delta_e-1)\Delta_e)}{1+L+\Delta_e-3L\Delta_e+2(L-2)\Delta_e^2}\\
\label{ex_im_im_2}
e^{-2\bar\kappa L}\frac{(e^{-\bar\kappa}-\Delta_e)^2}
{(e^{\bar\kappa}-\Delta_e)^2}=\frac{(1+e^{-\bar\kappa}-
2\Delta_e e^{-\bar\kappa})^2}{(1+e^{\bar\kappa}-2\Delta_e e^{\bar\kappa})^2}
\end{eqnarray}

Clearly, Eqs.\ \eref{ex_im_im}, \eref{ex_im_im_2} can be obtained from Eqs.\ \eref{ex_re_re},
\eref{ex_re_re_2} by redefining $\bar k\to i\bar\kappa$.

It is interesting to investigate the behavior of the exceptional points $\Delta_e^{(ii)}$ and
``amplitude'' $\bar\kappa$, solutions of~\eref{ex_im_im} \eref{ex_im_im_2}, as function of chain
size $L$ (Figure~\ref{fig_10}).  In the $L\to\infty$ limit, $\Delta_{e,1}^{(ii)}\to 1$ and
$\Delta_{e,2}^{(ii)}\to 1$.  Also, $\bar\kappa\to0$ for large chains.

\subsection{Fully edge-locked two-particle eigenfunctions} 
\label{edge_eig_2p}

\begin{figure}[t]
\begin{center}
\includegraphics[width=.99\textwidth]{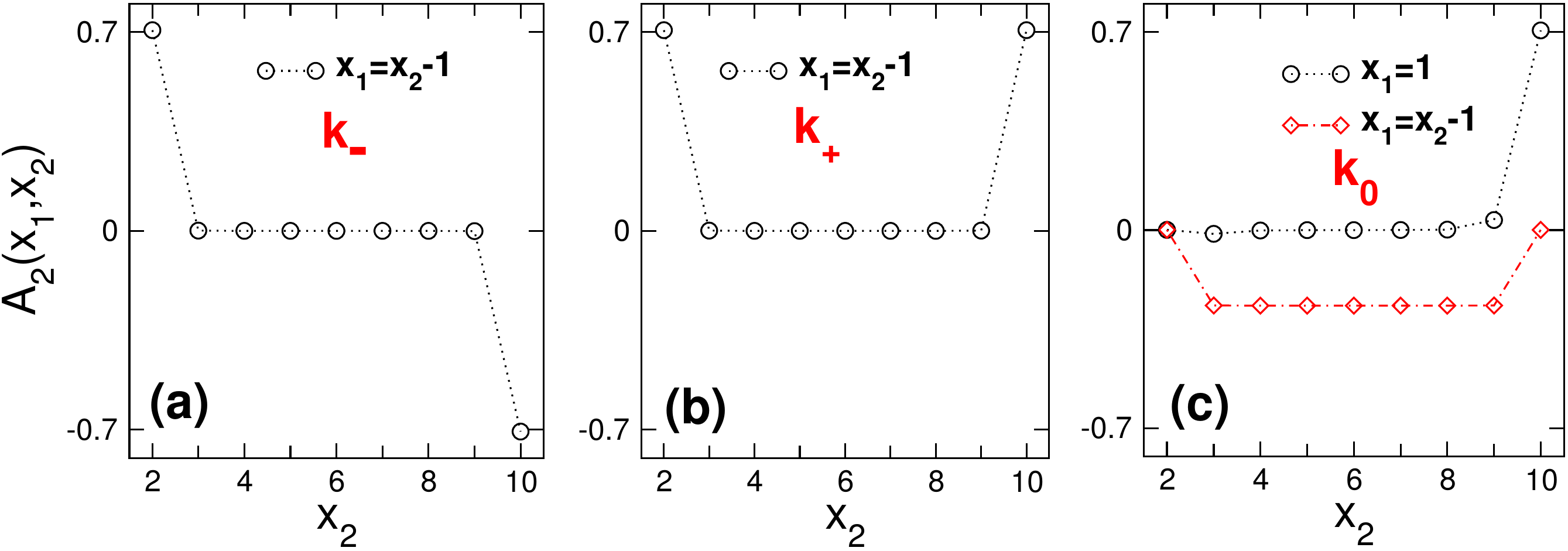}
\end{center}
\caption{ Edge-locked eigenfunctions corresponding to the \{Im,Im\} type solutions
  ($\mathbf{k}_{-}$, $\mathbf{k}_+$, $\mathbf{k}_0$). Data for $L=10$ chain at $\Delta=20$. The
  eigenfunction components $A_{2}(x_1,x_2)$ as function of $x_2$, the position of the second
  particle.  Only the components which are significantly nonzero are shown. (\textbf{a}) The
  eigenfunction obtained from the solution ${\mathbf k}_-$: $A_2(x_2-1,x_2)$ against
  $x_2$. (\textbf{b}) Same as in (\textbf{a}) but for the solution ${\mathbf k}_+$. (\textbf{c}) The
  eigenfunction amplitude obtained from ${\mathbf k}_0$:   $A_2(1,x_2)$
  (circles) and  $A_2(x_2-1,x_2)$ (rhombi).  In contrast to $\mathbf{k}_{\pm}$, this eigenfunction
  contains non-negligible bulk contributions, c.f.\ Eq.~\eref{eig_k0}.  }
\label{fig_11}
\end{figure}

In this section we discuss the edge-locked nature of the pure imaginary solutions of the Bethe
equations by constructing explicitly the corresponding eigenfunctions.  In Figure~\ref{fig_11} we
show for each of the three \{Im,Im\} solutions $\mathbf{k}_{\pm}$, $\mathbf{k}_0$ the corresponding
eigenvector components $A_2(x_1,x_2)$, i.e., the amplitudes of the configurations with particles at
positions $x_1<x_2$. Data are obtained by solving numerically the Bethe equations for a chain with
$L=10,\Delta=20$.  We consider only the configurations leading to significantly non zero amplitudes.
For instance one has that the eigenvectors $|\psi_\pm\rangle$ corresponding to ${\mathbf k}_\pm$ are
well approximated by

\begin{equation}
|\psi_{\pm}\rangle  ~\approx~ \frac{1}{\sqrt{2}}(|1,2\rangle\pm|L-1,L\rangle)
\end{equation}

with $|1,2\rangle,|L-1,L\rangle$ denoting the configurations with the 
two particles at positions $1,2$ and $L-1,L$ (respectively the first two and 
last two sites in the chain). Note that the two eigenfunctions correspond to 
states with opposite parity under ${\mathcal I}:i\to L-i+1$ (inversion with 
respect to the center of the chain).

The eigenvector obtained from ${\mathbf k}_0$ (cf. Figure~\ref{fig_11} (c)) is symmetric 
under ${\mathcal I}$ and is well approximated (in the large $\Delta$ regime) by 

\begin{equation}
\label{eig_k0}
|\psi_0\rangle ~\approx~ \frac{1}{\sqrt{2}} |1,L\rangle ~-~ \frac{1}{\sqrt{2(L-3)}}
\sum\limits_{m=2}^{L-2}|m,m+1\rangle
\end{equation}

Physically, $|\psi_0\rangle$ is a superposition of a pure edge-locked contribution and a bulk one.
The latter is an equal weight superposition of the configurations with the two particles next to
each other, i.e., a delocalized bound state.  The delocalized part is reminiscent of a string
solution with vanishing momentum (Section~\ref{string}).  The edge-locked part of the wavefunction
has on particle on either edge; there is no significant contribution from the configuration with
both particles on the same edge.

\section{Coexistence of extended and edge locked behavior: the \{Re,Im\} momentum pairs}
\label{re_im}

In this section we examine the $M=2$ solutions with one real and one imaginary Bethe momentum.
Clearly, the presence of the imaginary momentum signals edge-locking behavior for one particle,
while the real momentum signals that the other particle is extended (magnon-like).  The total number
of such eigenstates at large $\Delta$ is given as $2(L-3)$ (Table~\ref{table_1}) and corresponds to
the total number of configurations with only one particle localized at one edge of the chain and the
constraint that two particles cannot be on nearest-neighbor sites.

As $\Delta$ is lowered, exactly half of the \{Re,Im\} soutions turn into \{Re,Re\} solutions at
$L-3$ exceptional points.  The exceptional points are the same as those discussed in Section
\ref{sec_twoparticle_bothreal} ($\Delta^{(rr)}_{e,\ell}$; $\ell=1,2,\dots,L-3$) where $L-3$ extra
\{Re,Re\} solutions at small $\Delta$ vanish as $\Delta$ is increased.  In other words, the
vanishing \{Re,Re\} solutions observed in Figures \ref{fig_7} and \ref{fig_8} reappear as \{Re,Im\}
on the other (larger) side of $\Delta^{(rr)}_{e,\ell}$.  The location of the exceptional points and
the value of the other momentum at these points are therefore given by the nonlinear equations
\eref{ex_re_re}, \eref{ex_re_re_2}.

\begin{figure}[t]
\begin{center}
\includegraphics[width=.85\textwidth]{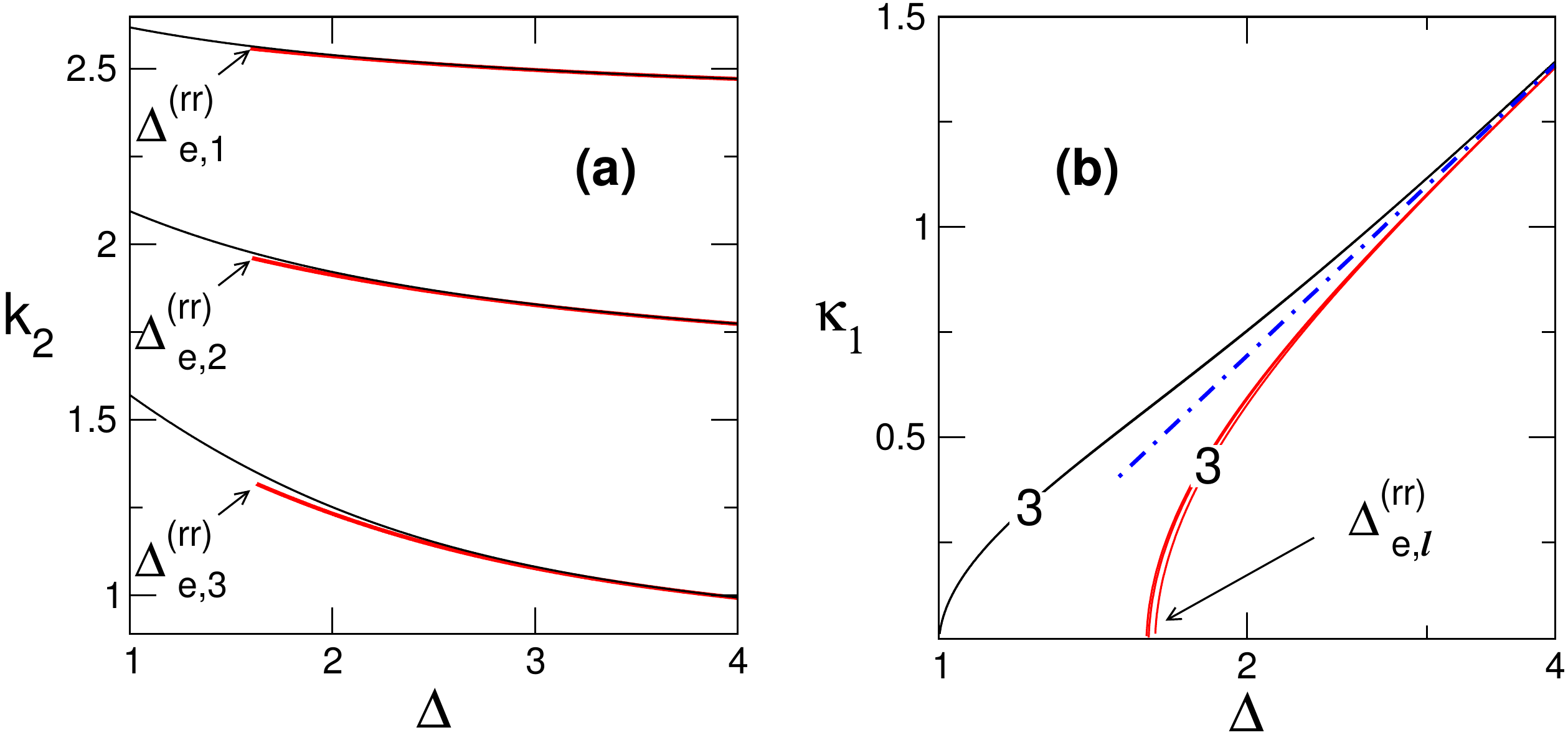}
\end{center}
\caption{ One real and one imaginary momentum, $(i\kappa_1,k_2)$ ($\kappa_1,k_2\in {\mathbb R}$), of
  solutions of \{Re,Im\} type.  Numerical solution of Bethe equations, for a $L=6$ chain,
  $1\le\Delta\le 4$.  There are $2(L-3)$ such solutions at large $\Delta$ and $(L-3)$ such solutions
  at small $\Delta$.  (\textbf{a}) The real momentum $k_2$ versus $\Delta$: the $L-3=3$ bands of
  solutions (two solutions per band). The arrows denote the exceptional points
  $\Delta_{e,\ell}^{(rr)}$ ($\ell=1,2,3$) at which $L-3$ solutions (one per band) disappear; these
  re-emerge as \{Re,Re\} solutions at smaller $\Delta$ below $\Delta_{e,\ell}^{(rr)}$.  The
  exceptional points are therefore the same as in Figure~\ref{fig_7}.  (\textbf{b}) The pure
  imaginary momentum $i\kappa_1$ of the pair.  The $\Delta$ axis is in logarithmic scale.  At the
  exceptional points $L-3$ pure imaginary momenta vanish.  The dashed-dotted line is the large
  $\Delta$ behavior, $\sim\log\Delta$.  }
\label{fig_12}
\end{figure}

This scenario is illustrated in Figure~\ref{fig_12} by plotting all the momentum pairs of type
$\{\textrm{Re},\textrm{Im}\}$ for a $M=2$, $L=6$, chain.  The imaginary and real components of the
generic pair are denoted as $\kappa_1$ and $k_2$.  While in the large $\Delta$ region there are
$2(L-3)=6$ momentum pairs, only half of them (i.e. $L-3$) are present near the Heisenberg point
$\Delta=1$. The other solutions disappear at the exceptional points $\Delta_{e,\ell}^{(rr)}$
($\ell=1,2,3$), which are the same as in Figure~\ref{fig_7}. It is worth mentioning that the $L-3$
states surviving across the exceptional points undergo the transformation
$\{\textrm{Re},\textrm{Im}\}\to\{\textrm{Re}, \textrm{Re}\}$ at $\Delta=1$, as can be surmised from
the vanishing of $\kappa_1$ at $\Delta\to 1$ in Figure~\ref{fig_12}(b).

\paragraph*{Bethe equations for  \{Re,Im\} solutions.}

We define $(k_1,k_2)\to(i\kappa_1,k_2)$, with $\kappa_1,k_2$ real.  The Bethe equations become
\begin{eqnarray}
\fl e^{-2(L+1)\kappa_1}\frac{(1-\Delta e^{\kappa_1})^2}{(1-
\Delta e^{-\kappa_1})^2}=\frac{(1+e^{ik_2-\kappa_1}-
2\Delta e^{-\kappa_1})(1+e^{ik_2+\kappa_1}-2\Delta 
e^{ik_2})}{(1+e^{ik_2-\kappa_1}-2\Delta e^{ik_2})
(1+e^{ik_2-\kappa_1}-2\Delta e^{\kappa_1})}\label{new2beteqna}
\\
\fl  e^{i2(L+1)k_2}\frac{(1-\Delta e^{-ik_2})^2}
{(1-\Delta e^{ik_2})^2}=\frac{(1+e^{ik_2-\kappa_1}-
2\Delta e^{ik_2})(1+e^{-ik_2-\kappa_1}-2\Delta 
e^{-\kappa_1})}{(1+e^{ik_2-\kappa_1}-2\Delta e^{-\kappa_1})
(1+e^{-ik_2-\kappa_1}-2\Delta e^{-ik_2})}
\label{new2beteqnb}
\end{eqnarray}
which, after redefining $z_1\equiv e^{\kappa_1}$, read   
\begin{eqnarray}
\fl\frac{(1-\Delta z_1)^2}{z_1^{2L}(z_1-\Delta)^2}
 =\frac{(1+e^{ik_2}/z_1-2\Delta/z_1 )(1+z_1e^{i
 k_2}-2\Delta e^{ik_2})}{(1+e^{ik_2}/z_1-
2\Delta e^{ik_2})(1+e^{ik_2}z_1-2\Delta z_1 )}
\label{reim_1}\\
\fl e^{i2(L+1)k_2}\frac{(1-\Delta e^{-ik_2})^2}
{(1-\Delta e^{ik_2})^2}=\frac{(1+e^{ik_2}/z_1-2
\Delta e^{ik_2})(1+e^{-ik_2}/z_1-2\Delta/z_1)}
{(1+e^{ik_2}/z_1-2\Delta/z_1)(1+e^{-ik_2}/z_1-2
\Delta e^{-ik_2})}\label{reim_2}
\end{eqnarray}
We focus on the denominator of the left side in~\eref{reim_1}, considering large $\Delta$ and $L$.
Using $z_1>1$ one has that the term $z_1^{2L}$ grows exponentially with $L$.  The right side
of~\eref{reim_1} shows a different behavior (it vanishes mildly with increasing $L$) unless
$z_1=\Delta+e^{-\beta L}$ with $\beta>0$. After fixing $z_1=\Delta$ the degree of freedom of one
particle is ``frozen'' (the particle is edge locked) and the Bethe like equation for the momentum
$k_2$ of the ``free'' (i.e. not locked) one is given as

\begin{equation}
\label{im_re_be}
e^{i2(L+1)k_2}=\frac{(1+e^{ik_2}/\Delta
-2\Delta e^{ik_2})(1-e^{-ik_2}/\Delta)(1-
\Delta e^{ik_2})^2}{(1+e^{-ik_2}/\Delta-
2\Delta e^{-ik_2})(1-e^{ik_2}/\Delta)(1-
\Delta e^{-ik_2})^2}
\end{equation}

Taking the logarithm of both sides of~\eref{im_re_be} one has

\begin{eqnarray}
\label{log_reim}
\fl (L+1)k_2 ~=~ \pi J_\alpha+\arctan\frac{\sin k_2}
{\Delta-\cos k_2} ~-
\\ \nonumber
\qquad 2\arctan\frac{\Delta\sin k_2}
{1-\Delta\cos k_2}+\arctan\frac{(1-2\Delta^2)
\sin k_2}{\Delta+(1-2\Delta^2)\cos k_2}
\end{eqnarray}

Numerically, we find that Eq.\ \eref{log_reim} admits $(L-3)$ real solutions for $k_2$, which are
obtained choosing the Bethe numbers as $J_\alpha=\{1,2,\dots,\lfloor (L-3)/2\rfloor\}\cup
\{L-\lceil(L-3)/2\rceil+1, \dots,L\}$.  Each solution of~\eref{log_reim} has to be counted with a
degeneracy factor two, in order to recover the correct counting $2(L-3)$.  In fact, including the
higher order corrections in~\eref{reim_1}, i.e., going beyond the approximation $z_1\sim\Delta$, has
the effect of splitting in two every solution of~\eref{log_reim}.  The splitting vanishes
exponentially with $L$, implying that for large enough $L$ the solutions of~\eref{log_reim} (with
$\kappa_1=\log\Delta$) coincide, in practice, with the solutions of~\eref{reim_1}, \eref{reim_2}.

\section{Bound state of two delocalized particles: string solutions}
\label{string}

We now discuss the solutions of the Bethe equations which are complex conjugate pairs (``strings''),
which describe eigenstates where the two particles are mutually bound and spatially extended.  While
other classes of solutions (cf. Table~\ref{table_1}) change nature upon changing the anisotropy
$\Delta$, it turns out that all the string solutions are stable in the whole region
$\Delta\in(1,\infty)$, i.e., there is no unbinding of the pair at any $\Delta>1$.

\begin{figure}[t]
\begin{center}
\includegraphics[width=.99\textwidth]{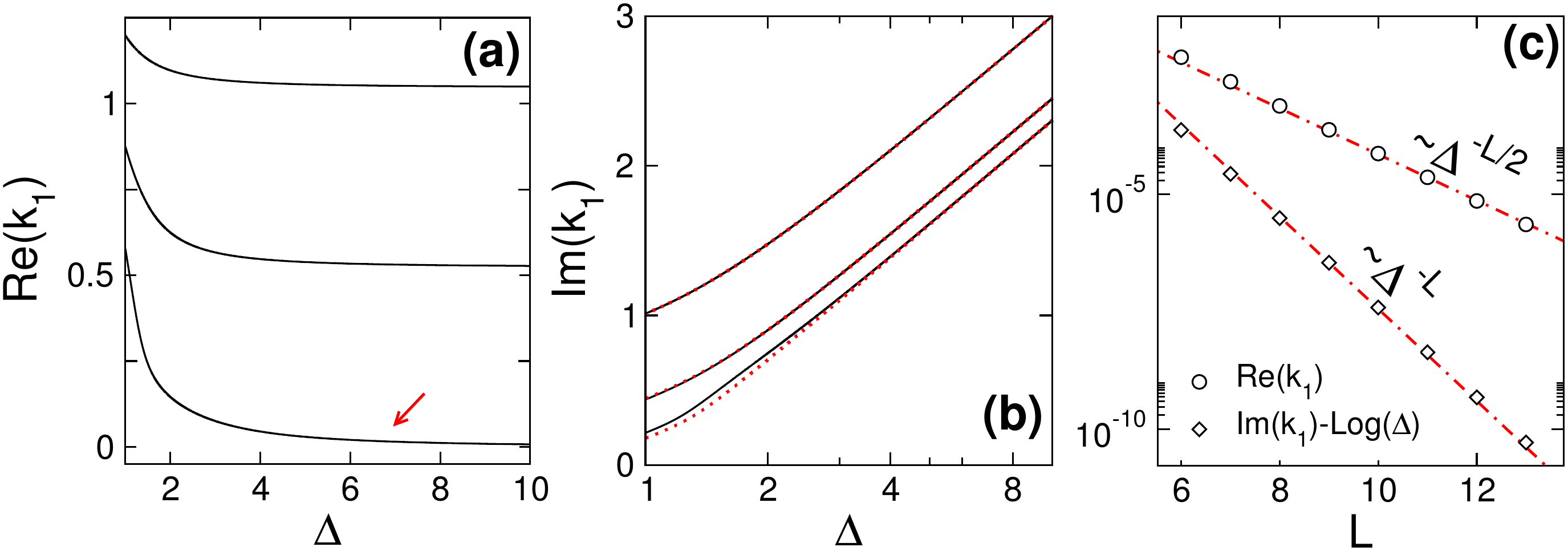}
\end{center}
\caption{ String solutions $(k_2 \equiv k_1^*)$ representing bound delocalized states.  Data for a
  $L=6$ chain. Note that the number of string solutions ($L-3$) does not change with $\Delta$ (no
  exceptional points).  (\textbf{a}) Real part of $k_1$ as a function of $\Delta$. The
  $\Delta\to\infty$ asymptotic values are described by~\eref{log_bt}.  The lowest solution (marked
  by the arrow) shows vanishing momentum.  (\textbf{b}) Imaginary part of $k_1$ as function of
  $\Delta$. The asymptotic behavior is given as $\log[\Delta/\cos(\textrm{Re}(k_1))]$ (dotted
  lines). (\textbf{c}) The string with vanishing real part (c.f.\ panel (\textbf{a})):
  $\textrm{Re}\, (k_1)$ and $\textrm{Im}\, (k_1)-\log\Delta$ as function of the chain size $L$ at
  fixed $\Delta=10$. Dashed-dotted lines are fits to $\sim\Delta^{-L/2}$ and $\sim\Delta^{-L}$.  }
\label{fig_13s}
\end{figure}

After defining $k_1=a+i b$ and $k_2=a-i b$, and using the parametrization  $z\equiv e^{b}$, the Bethe equations read
%
%
\begin{eqnarray}
\fl  e^{i2(L+1)a}/z^{2(L+1)}=\frac{(1-\Delta e^{ia}/z)^2(1+e^{2ia}-
2\Delta e^{ia}/z)(1+z^2-2\Delta e^{ia}z)}{(1-\Delta e^{-ia}z)^2
(1+e^{2ia}-2\Delta e^{ia}z)(1+z^2-2\Delta e^{-ia}z)} \label{string1a}\\
\fl e^{i2(L+1)a}z^{2(L+1)}=\frac{(1-\Delta e^{ia}z)^2(1+e^{2ia}-
2\Delta e^{ia}z)(1+1/z^2-2\Delta e^{ia}/z)}{(1-\Delta e^{-ia}/z)^2
(1+e^{2ia}-2\Delta e^{ia}/z)(1+1/z^2-2\Delta e^{-ia}/z)}\label{string2a}
\end{eqnarray}
which can be rewritten as 
\begin{eqnarray}
\label{eq_trig_1}
\frac{(z\cos a-\Delta)(1+z^2-2z\Delta e^{ia})}{z(z\Delta-\cos a)
(1+z^2-2z\Delta e^{-ia})}+z^{-2L}e^{2iLa}\frac{(1-z\Delta e^{-ia})^2}
{(\Delta -ze^{-ia})^2}=0\\
\label{eq_trig_2}
\frac{z(z\Delta-\cos a)(1+z^2-2z\Delta e^{ia})}{(z\cos a-\Delta)
(1+z^2-2z\Delta e^{-ia})}+z^{2L}e^{2iLa}\frac{(\Delta -ze^{ia})^2}
{(1-z\Delta e^{ia})^2}=0
\end{eqnarray}

We observe that the second term in equation \eref{eq_trig_1}, [\eref{eq_trig_2}] vanishes [diverges]
exponentially with $L\to\infty$ if the imaginary part of the momentum (i.e. $b$) is negative
[positive]. In order to match the behavior of the two terms one can impose $z\sim\Delta/\cos a$,
neglecting (additive) exponentially vanishing contributions in the limit $L\to\infty$. After fixing
$z=\Delta/\cos a$ in~\eref{eq_trig_1}, \eref{eq_trig_2}, and multiplying the two Bethe equations to
cancel the vanishing denominator in~\eref{eq_trig_2} one obtains

\begin{equation}
\fl\frac{(1+e^{2ia}-2\Delta e^{2ia})^2(1+e^{2ia}+2\Delta e^{2ia})^2}
{(1+e^{-2ia}-2\Delta e^{-2ia})^2(1+e^{-2ia}+2\Delta e^{-2ia})^2}=
e^{4ia(L+3)}\frac{(1+e^{-2ia}-2\Delta^2e^{-2ia})^2}{(1+e^{2ia}-
2\Delta^2e^{2ia})^2}
\end{equation}

or in logarithmic form 

\begin{eqnarray}
\label{log_bt}
\fl (L+3)a=\pi J+\arctan\frac{(1-2\Delta^2)\sin(2a)}{1+(1-2\Delta^2)
\cos(2a)}+\\\nonumber\qquad\arctan\frac{(1-2\Delta)\sin(2a)}{1+(1-2\Delta)
\cos(2a)}+\arctan\frac{(1+2\Delta)\sin(2a)}{1+(1+2\Delta)\cos(2a)}
\end{eqnarray}

The equation above is similar to the so-called Bethe-Takahashi 
equations~\cite{taka-1972,ilakovac-1999,fujita-2003,hagemans-2007} that appear in the study of string solutions 
in the XXZ chain with periodic boundary conditions.

The quantum numbers $J\in[1,L]$ identify the different solutions of \eref{log_bt}.  Note that there
is a subtlety in equation~\eref{log_bt}: the number of real solutions is found numerically to be
$L-4$.  To get the correct number of strings ($L-3$) from these leading-order large-$L$ equations,
one should include the solution $a=0$ (which implies $z\to\log\Delta$).  Higher order corrections to
$z$, going beyond $z\sim\Delta/\cos a$, would make this solution non vanishing.

Interestingly, this implies that the string solution with $a\to 0$ is almost degenerate in energy
with one of the edge-locked states, namely, the edge-locked eigenstate obtained from the momentum
pair ${\bf k}_0$ of Figure~\ref{fig_9}.  Figure \ref{fig_13s}(a) and Figure \ref{2p_spectrum}
highlight this string solution with red arrows.

All these findings are supported numerically in Figure~\ref{fig_13s} plotting 
the momenta $k_1$ $k_2$ for all the string solutions for a 
chain with $L=6$ as function of the anisotropy $1<\Delta<10$ (plot of $\textrm{Re}(k_1)=a$ 
and $\textrm{Im}(k_1)=b$, respectively panels (a) and (b) in Figure~\ref{fig_13s}). 
Data are obtaining by solving numerically the Bethe equations. In Figure~\ref{fig_13s} 
(a) the arrow denotes the string solution with vanishing real part. Figure~\ref{fig_13s} 
(b) shows the imaginary parts of the string solutions (continuous lines is  
$\textrm{Im}(k_1)$ versus $\Delta$). The dotted lines is $\log(\Delta/\cos\textrm{Re}( 
k_1))$, where for $\textrm{Re}(k_1)$ the values reported in panel (a) are used.  

The difference between $\textrm{Im}(k_1)$ and $\log(\Delta/\cos\textrm{Re}(k_1))$ is not 
visible, except for the lowest curve, especially at large $\Delta$. More information about the 
string with vanishing momentum is shown in panel (c) plotting both $\textrm{Re}(k_1)$ and 
$\textrm{Im}(k_1)-\log\Delta$ versus $L$ at fixed $\Delta=10$ (logarithmic scale on 
the $y$-axis). Clearly deviations from the asymptotic (i.e. at large $L$) solution 
$\textrm{Im}(k_1)=i\log\Delta, \textrm{Re}(k_1)=0$ decay exponentially with increasing $L$. 

\section{The two-particle energy spectrum: edge-locking vs extended behavior}
\label{sec_2p_spectrum}

\begin{figure}[t]
\begin{center}
\includegraphics[width=.75\textwidth]{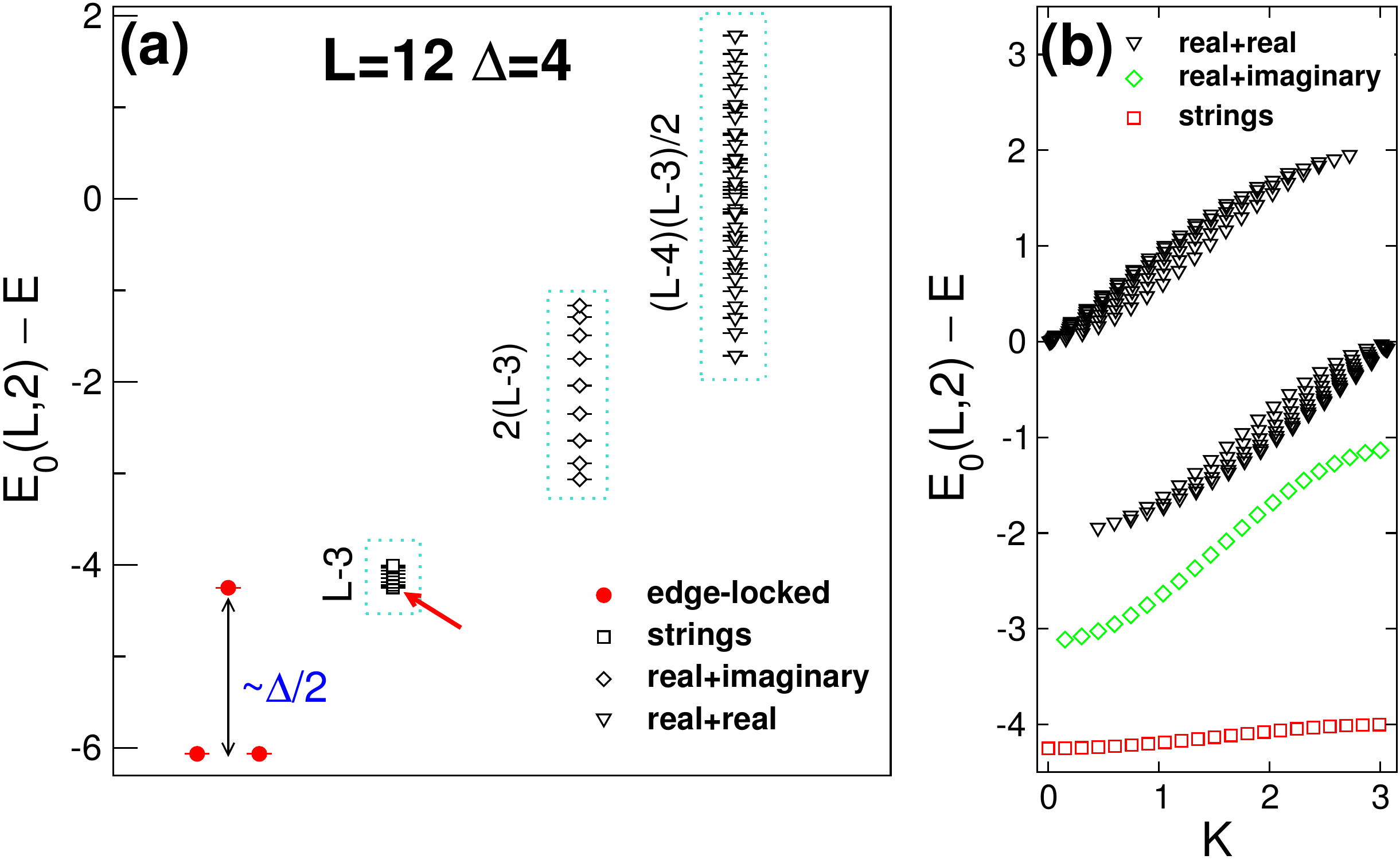}
\end{center}
\caption{ Energy spectrum of the open XXZ spin chain in the sector with two particles, inverted and
  added to $E_0(L,2)\equiv \frac{1}{4}(L-9)\Delta$.  Data shown for a $L=12$ chain at $\Delta=4$.
  (\textbf{a}) The four types of eigenstates are separated horizontally, the $x$-axis is otherwise a
  dummy axis.  The lowest three levels are the edge-locked states (\{Im,Im\} solutions). The total
  number of energy levels for each type of solution is also shown. The lowest energy level in the
  string sector (arrow) is almost degenerate with one of the edge-locked states. The levels of
  \{Re,Im\} type are almost doubly degenerate. (\textbf{b}) Dispersion of the different types of
  eigenstates, plotted as the energy versus the total real momentum $K\equiv
  \mathrm{Re}(k_1)+\mathrm{Re}(k_2)\, [\mathrm{mod}\,\pi]$.  }
\label{2p_spectrum}
\end{figure}

In this section we analyze the full two-particle energy spectrum of the open-boundary XXZ spin
chain, at anisotropies larger than the region of exceptional points.  In particular, for each class
of solution (\{Re,Re\}, \{Re,Im\}, etc) of the Bethe equations we isolate the corresponding
contributions to the spectrum.  In Figure~\ref{2p_spectrum} we show the spectrum for an $L=12$ chain
at $\Delta=4$, obtained from numerically solving the Bethe equations.  We choose to plot the
quantity $E_0(L,2)-E$, i.e., the energies are inverted and added to $E_0(L,2)\equiv
\frac{1}{4}(L-9)\Delta$, so that the fully edge-locked states appear at the bottom.

The lowest three levels (full circles in the Figure) correspond to the edge-locked states.  The
doublet at the bottom and the isolated level above are given respectively by the half-string
solutions ${\mathbf k}_\pm$ and ${\mathbf k}_0$ (cf. Figure~\ref{fig_9}). The leading large-$\Delta$
behaviors are obtained from~\eref{n_deg_theo} and~\eref{large_L_deg} as
\begin{equation}
E_0(L,2)-E({\mathbf k}_\pm)\sim-\frac{\Delta-3\Delta^3}{1-2\Delta^2}\sim
-\frac{3}{2}\Delta
\end{equation}
and
\begin{equation}
E_0(L,2)-E({\mathbf k}_0)\sim-\frac{\Delta^4-\Delta^L\sqrt{L-3}(1+\Delta^2)}
{\Delta^3-\sqrt{L-3}\Delta^{1+L}}\sim -\Delta
\end{equation}
The lowest dispersing levels above the edge-locked sector are the strings.  The lowest level in the
string sector, which corresponds to vanishing total Bethe momentum, is degenerate at large $\Delta$
with the edge-locked level obtained from the solution ${\bf k}_0$.  Higher levels in the spectrum
correspond to the \{Re,Im\} type and the two magnons (type \{Re,Re\}), which contribute respectively
with $2(L-3)$ and $(L-4)(L-3)/2$ energy levels.

In Figure~\ref{2p_spectrum}(b), $E_0(L,2)-E$ is plotted against the total real part of Bethe
momenta, $K\equiv \mathrm{Re}(k_1)+\mathrm{Re}(k_2)\, [\mathrm{mod}\,\pi]$.  For the \{Re,Re\} type
solutions, $k_1+k_2$ ranges from 0 to $2\pi$, and therefore the \{Re,Re\} band appears split into
two in the $(0,\pi)$ domain.

The width of the dispersion in the string states is smaller compared to the other classes of states
and depends on the anisotropy $\Delta$.  The dispersion of the string states, using the results in
section~\ref{string}, is obtained at leading order in $1/\Delta$ to be
\begin{equation}
E_0(L,2)-E\sim-\Delta -\frac{\cos^2 (K/2)}{\Delta}
\end{equation}
The width $\sim 1/\Delta$ vanishes in the Ising limit $\Delta\to\infty$.  Physically, this is
because two-particle bound states are ``heavy'' objects with effective hopping strength $\sim
1/\Delta$ \cite{haque-10, Ganahl_PRL2012}.  The dependence of the dispersion with $\Delta$ is
different for the other two classes of dispersive solutions (\{Re,Re\} and \{Re,Im\}), for which the
dispersion width does not change significantly with $\Delta$ and $L$, and is given respectively by
$\delta E\sim 4$ and $\delta E\sim 2$.

\section{Conclusions: Summary and Perspectives}
\label{conclusions}

\paragraph*{Summary.}

In this Article, we investigated edge-locking behavior in the eigenstates of the spin-$\frac{1}{2}$
Heisenberg XXZ chain with open boundary conditions, in the highly polarized sectors $M=1$ and $M=2$.
Exploiting the Bethe ansatz solution of the model we constructed explicitly the full spectrum
(energies and eigenfunctions), focusing on the region at $\Delta>1$.  We presented a complete
classification of all the possible solutions of the Bethe equations in the whole region
$\Delta\in(1,\infty)$.

Edge-locked eigenstates are those in which one or more of the particles are exponentially localized
at the edges of the chain.  In all sectors (e.g., the $M=1$ and $M=2$ sectors we have detailed),
there are two eigenstates where the particles are all localized at the left or at the right edge.
In addition, for $M>1$ we can have some particles localized at the left edge and some localized at
the right edge.  These fully edge-locked eigenstates are all associated with pure imaginary solutions
of the Bethe equations.  In contrast, real solutions of the Bethe equations reflect extended
(i.e.\ ``magnon''-like) behavior.  For $M>1$, we naturally can have eigenstates where some of the
particles are extended and some are edge-localized, i.e., solutions with some momenta real and some
momenta imaginary.  In the $M=2$ case, these show up as \{Re,Im\} type of solutions.  Finally, for
$M>1$ one can also have string solutions with the particles bound but delocalized.  

At large $\Delta$, one can use combinatorial counting of different spatial configurations and find
out the numbers of eigenstates of different types, i.e., the numbers of real and imaginary solutions
in the $M=1$ case (Section \ref{one_p}), and the numbers of \{Re,Re\}, \{Im,Im\}, \{Re,Im\}, and
string types of solutions in the $M=2$ case (Section \ref{two_part}).  

At any finite chain length $L$, as one decreases $\Delta$, we find that there are special values of
$\Delta$ where some of the imaginary momenta pass through zero and become real momenta.  In the
$M=1$ case, one of the two imaginary solutions becomes real at $\Delta_e=\frac{L+1}{L-1}$, so that
we are left with a single imaginary solution, i.e., a single edge-locked eigenstate, at
$\Delta\in(1,\Delta_e)$.  For $M=2$, we have transitions from \{Im,Im\} to \{Re,Im\}, or from
\{Re,Im\} to \{Re,Re\}, as $\Delta$ is lowered.  This corresponds to some of edge-locked particles
getting delocalized and becoming extended in the bulk of the chain.  We have characterized in detail
these changes of eigenstates, the positions of the ``exceptional points'' where they occur, and the
behavior of the Bethe momenta near these points and at large $\Delta$.  In the large-chain
($L\to\infty$) limit, the exceptional points all coalesce at the isotropic point $\Delta=1$, so that
all the edge-locking present at large $\Delta$ become stable in the whole $\Delta\in(1,\infty)$
region.  The string solutions (complex conjugate momentum pairs) are found to be stable with no
change of character in the whole $\Delta\in(1,\infty)$ region, even at finite $L$.

We have also presented spectral signatures of edge-locking (Figures
\ref{fig_spectra_periodic_vs_open}, \ref{fig_3}, \ref{2p_spectrum}).  At large $\Delta$, in all $M$
sectors, two of the fully edge-locked states are distinctly well-separated from the rest of the
spectrum, by a gap $\sim\Delta$.  The energy splitting within this doublet vanishes exponentially
with the chain length.  In the $M=2$ sector, there is an additional edge-locked state which has a
$\ket{1,L}$ component (one particle edge-locked at each edge), and also a string-like component
extended in the bulk, c.f., Eq.~\eref{eig_k0}.  This new fully edge-locked state is the third state
at the bottom of the inverted spectrum and nearly degenerate with the edge of the band of string
eigenstates (Figure \ref{2p_spectrum}).

\paragraph*{Open Issues.}

The present work opens up a number of research avenues.  

We have focused on the region $\Delta\in(1,\infty)$, where we have detailed the different types of
eigenstates and related them to edge-locking, and shown how there are transformations between
different types as a function of $\Delta$ and $L$.  The intuition at $\Delta\gg{1}$, where spin
configuration counting gives accurate classification and counting of the different types of
eigenstates, has been particularly helpful.  Clearly, the situation should be quite different at the
isotropic point, $\Delta=1$, and at smaller anisotropies, $\Delta<1$.  Classifying the eigenstates
according to their edge behavior for the open XXZ chain in these regimes remains an open task.

We have restricted ourselves to the $M=1$ and $M=2$ sectors, since there was substantial detail to be
worked out in these cases.  It is known that the $M>2$ sectors contain richer edge-related behavior
and novel types of edge-locking phenomena \cite{haque-10}.  In particular, for larger $M$, there is
a hierarchy of locking behaviors at increasing distances from the edge, related to sub-structures
with smaller gaps in the energy spectrum.  A Bethe ansatz description of $M>2$ sectors for the
open-boundary XXZ chain is thus expected to include a rich set of behaviors beyond those explored in
the present study.



\end{document}